\documentclass[aps,prb,twocolumn,groupedaddress,showpacs,superscriptaddress,amssymb,amsmath]{revtex4-2}
\pdfoutput=1 
\usepackage{physics}

\usepackage{slashed}

\newcommand{\be}{\begin{equation}}
\newcommand{\ee}{\end{equation}}

\usepackage{subcaption}
\usepackage{ucs}
\usepackage[utf8x]{inputenc}
\usepackage{amsmath,bm}
\usepackage{amsfonts}
\usepackage{comment}
\usepackage{amssymb}
\usepackage{array}
\usepackage{epsfig}
\usepackage{slashed}
\usepackage{lipsum}
\usepackage[colorlinks, linkcolor=red,citecolor=blue,urlcolor=blue]{hyperref}

\begin{document}
\title{Quantum Regression theorem for multi-time correlators : A detailed analysis in the Heisenberg Picture}
\author{Sakil Khan}
\email{sakil.khan@students.iiserpune.ac.in}
\affiliation{Department of Physics,
		Indian Institute of Science Education and Research, Pune 411008, India}
\author{Bijay Kumar Agarwalla}
\email{bijay@iiserpune.ac.in}
\affiliation{Department of Physics,
		Indian Institute of Science Education and Research, Pune 411008, India}
\author{Sachin Jain}
\email{sachin.jain@iiserpune.ac.in}
\affiliation{Department of Physics,
		Indian Institute of Science Education and Research, Pune 411008, India}

\date{\today}

\begin{abstract}
Quantum regression theorem is a very useful result in open quantum system and extensively used for computing multi-point correlation functions. Traditionally it is derived for two-time correlators in the Markovian limit employing the Schr\"odinger picture. In this paper we make use of the Heisenberg picture to derive quantum regression theorems for multi-time correlation functions which in the special limit reduce to the well known two-time regression theorem. For multi-time correlation function we find that the regression theorem takes the same form as it takes for two-time correlation function with a mild restriction that one of the times should be greater than all the other time variables. Interestingly, the Heisenberg picture also allows us to derive analogue of regression theorem for out-of-time-ordered correlators (OTOCs). We further extend our study for the case of non-Markovian dynamics and report the modifications to the standard quantum regression theorem. We illustrate all of the above results using the paradigmatic dissipative spin-boson model. 
\end{abstract}
  \maketitle  

\section{Introduction}
 Correlation functions are important dynamical quantities which are often related to experimentally measurable quantities. In the context of open quantum system \cite{Smirne2021,davies,Alicki,Carmichael, qn,breuer} with the system of interest following  a Markovian dynamics, Quantum Regression Theorem (QRT)\cite{Carmichael, qn,breuer,qrt} turns out to be one of the most useful and practical tools to compute the correlation functions \cite{Carmichael}. However, it is limited to certain special kind of time arrangements and thus allowing to compute only special types of correlation functions. The QRT states that the knowledge of time evolution of a single-point function is sufficient to determine the time evolution of two-point or multi-point  correlation functions.
%
%
More explicitly, the validity of QRT requires that there exists a complete set of system operators $A_{\mu}, \mu=1,2,...$ such that 
  \begin{equation}\label{coma}
     \frac{d}{dt} \langle A_{\mu}(t)\rangle =\sum_{\lambda} M_{\mu \lambda}\langle A_{\lambda}(t)\rangle.
  \end{equation}
  Then the QRT for two and three-point function reads as,
  \begin{align}\label{eq8'}
   \frac{d}{d\tau}\langle O(t)A_{\mu}(t+\tau)\rangle&\!=\!
  \sum_{\lambda} M_{\mu \lambda}\langle O(t)A_{\lambda}(t+\tau)\rangle,\nonumber\\
  \frac{d}{d\tau}\langle O_{1}(t)A_{\mu}(t+\tau)O_{2}(t)\rangle&\!=\!\sum_{\lambda}\, M_{\mu \lambda}\langle O_{1}(t)A_{\lambda}(t+\tau)O_{2}(t)\rangle, \nonumber\\
  \frac{d}{d\tau} \langle A_{\mu}(t+\tau)O_{1}(t) O_{3}(t)\rangle &\!=\!\sum_{\lambda}\, M_{\mu \lambda} \langle A_{\lambda}(t+\tau)O_{1}(t) O_{3}(t)\rangle.
  \end{align}
Interestingly, it is easy to generalize the QRT in Eq.~\eqref{eq8'} for arbitrary $N$-point correlation functions of the form 
\be 
 \label{eqnpt}\langle A_1(t) A_2(t + \tau) A_3 (t) \cdots A_n(t)\rangle.
\ee 
where the position of the operator with argument $t+\tau$ can be arbitrary. {\textcolor{black}{Note that the above QRTs are given for multi-point correlation functions which are dependent on two-times $t$ and $\tau$. }}
Recently there has been a lot of research activity to understand systems that follow non-Markovian dynamics \cite{nv1,nmv2,nmv3,nmv4,nmv5} and attempt has been made to show violation of regression theorems for such systems \cite{nmr1,PhysRevA.75.052108,nmvr3,nmvqr7,nmv10,nmv11,nmv12}. Inspite of its great importance and interest, there has been a lack of systematic derivation of Regression type theorem for cases beyond two-time correlation function and its extension for non-Markovian systems. 
In general, regression theorem may not hold for general time configurations such as 
\begin{equation}
    \langle A(t_1) B(t_2) C(t_3)\rangle
\end{equation}
or more generally
\begin{equation}\label{gnc}
    \langle A_1(t_1) A_2(t_2)\cdots A_n(t_n)\rangle.
\end{equation}
 Here we would like to understand if there exist QRT type relations for such class of correlation functions including the out-of-time-ordered correlators (OTOCs) \footnote{OTOC is defined as $\langle O_{1}(t_{1})O_{2}(t_{2})O_{3}(t_{1})O_{4}(t_{2})\rangle. $ Even though it's a two-time correlator it's form significantly differ from equation  \eqref{eqnpt}.} which are a special class of correlation functions \cite{otoc1}. 

As in the Schr\"odinger picture it is difficult to define multi-time correlators of the above form, it is therefore more appropriate and suitable to consider such correlators in the Heisenberg picture \cite{karve2020heisenberg}. We therefore, in this paper, derive regression theorem for multi-time correlators, in the Markovian limit, using the Heisenberg picture, as discussed in Ref.~\cite{karve2020heisenberg}. We also extend our study for the non-Markovian systems as well. 

The paper is organised as follows:  In section \ref{QRTh1}, we first start with deriving the QRT for two-point functions and then extend our analysis  to three, four and further generalize to $N$-point functions with general time arrangements. We also point out for what special time arrangements the QRT may not hold. In section \ref{OTOCs}, we derive a regression like expression for the OTOC. In section \ref{exds}, we illustrate all these above findings for a paradigmatic dissipative spin-boson model. 
In section \ref{nmv}, we  give a systematic derivation of the equation similar to QRT for non-Markovian systems. We skip most of the lengthy derivations to the appendix to keep the discussion in the main text transparent.

\section{QRT using the Heisenberg picture}\label{QRTh1}
In this section we derive QRT for multi-time correlation functions. We make extensive use of the Heisenberg picture formulation for open quantum system, recently described in \cite{karve2020heisenberg} and for completeness we also review this formalism in Appendix \ref{Heisencor} and further discuss the Markovian limit in Appendix \ref{markov1}. Our first aim here is to derive the well-known  forms of QRTs for two-point and special three-point functions, as discussed in the introduction, using the Heisenberg picture. We then aim for generalizing the QRTs  for more generic $N$- point correlation functions defined with multi-time arguments. We also discuss the limitations of the QRT in the Markovian limit and also generalize our study to  non-Markovian systems.
  
Let the start by writing the Hamiltonian of the total sy  $H  = H_{S}+H_{R}+\lambda H_{S R}$, where $H_S$ is the Hamiltonian of the system of interest, $H_R$ represents the Hamiltonian for the bath (reservoir), and $H_{SR}$ is the coupling Hamiltonian between the system and the bath. We keep the parameter $\lambda$ to keep track of the order of the perturbation with respect to system-bath interaction. We also make the standard choice for the initial condition of the total density operator at $t=0$ by considering a product initial state between the system and the bath and write $\rho_{SR}(t=0) =\rho_{S} \otimes \rho_{R}$.  The interaction between the system and the bath is turned on at $t=0^{+}$. Since in the Heisenberg picture, the operators evolve in time, it is important to define the reduced system operators. The one point reduced operator is defined as  $O_{S}(t)={\rm Tr}_{R}[O(t)\rho_{R}]$ where the operator $O(t)$ evolves unitarily with respect to the full Hamiltonian $H$.  The expectation value of the operator $O$ at time $t$ then can be written as,
 \begin{equation}
      \langle O(t)\rangle={\rm Tr}_{S}[{\rm Tr}_{R}[O(t)\rho_{R}]\rho_{S}]={\rm Tr}_{S}[O_{S}(t)\rho_{S}].
  \end{equation}
In a similar manner, one can define arbitrary $N$-point reduced operator as
 \begin{eqnarray}
 \label{eqredo}
 &&(O_{1}(t_{1})O_{2}(t_{2})....O_{N}(t_{N}))_{S}= \nonumber \\
  && \quad \quad {\rm Tr}_{R}[O_{1}(t_{1})O_{2}(t_{2})....O_{N}(t_{N})\rho_{R}].
 \end{eqnarray}
This definition of reduced operator has a property that  $(O_{1}(t_{1})O_{2}(t_{2}) \cdots O_{N}(t_{N}))_{S} \neq O_{1 S}(t_{1})O_{2 S}(t_{2}) \cdots O_{N}(t_{N})$ as a result of finite system-bath coupling.
  
 To derive the QRT in the Heisenberg picture we write an analogous equation like equation \eqref{coma} in the Heisenberg picture by assuming that there exists a complete set of  reduced system operators $A_{\mu S}(t)$ that satisfies \footnote{ In one of the examples discussed later, we explicitly construct operator $A_\mu$} the relation
 \begin{equation}\label{eqe}
     \frac{d}{dt} A_{\mu  S}(t)=\sum_{\lambda} M_{\mu  \lambda}A_{\lambda  S}(t).
  \end{equation} 
  The above equation implies that the operators form a  closed set  between themselves.

 \subsection{QRT for two-point correlation functions}
 To keep our discussion simple, we first focus on deriving the QRT for two-point correlation functions. Following the definition in \eqref{eqredo} for two-point reduced operators, one can write \cite{karve2020heisenberg}
 (please see Appendix \ref{Heisencor} for details of the derivation) 
  \begin{equation}\label{eq2pt1}
    ( O_{1}(t_{1})O_{2}(t_{2}))_{S}=O_{1  S}(t_{1})O_{2  S}(t_{2})+
 {I}[O_{1  S}(t_{1}),O_{2  S}(t_{2})] 
  \end{equation}
  where $O_{1  S}(t_{1}),O_{2  S}(t_{2})$ are reduced one-point operators and  $I[O_{1  S}(t_{1}),O_{2  S}(t_{2})] $ is called the irreducible part capturing the information about coupled system-bath dynamics. \footnote{See equation (\ref{eqf}) and (\ref{eq26'})-(\ref{eq28'}) in the appendix \ref{markov1} for details.} 
  
One can explicitly calculate the quantity $I$ upto the second-order of the system-bath coupling ($\lambda^2$) in both Markovian and non-Markovian limits. 
Given the above expression, the two-point correlation functions can be easily computed by performing an additional trace over the initial system density operator i.e.,
 \begin{equation}\label{eqm}
      \langle O_{1}(t_{1})O_{2}(t_{2})\rangle={\rm Tr}_{S}[(O_{1}(t_{1})O_{2}(t_{2}))_{S}\rho_{S}(0)]
  \end{equation}
 To derive the QRT, we set, $O_{2}=A_{\mu},     O_{1}=O,$ and  consider $t_{2}>t_{1}$. Now taking derivative with respect to $t_{2}$ in equation \eqref{eq2pt1}, the first term of the of the right hand side immediately gives a QRT like expression
\begin{equation}\label{eq14}
    \frac{d}{dt_{2}} (O_{S}(t_{1})A_{\mu  S}(t_{2})) =\sum_{\lambda} M_{\mu \lambda} (O_{S}(t_{1})A_{\lambda  S}(t_{2})).
\end{equation}
thanks to Eq.~\eqref{eqe}. Now interestingly, one can show that for $t_2 > t_1$ and in the Markovian limit, the irreducible part $I[O_{S}(t_{1}),A_{\mu  S}(t_{2})]$ also satisfies a regression like expression, given as,
\begin{equation}
    \frac{d}{dt_{2}}\label{eqg} I[O_{S}(t_{1}),A_{\mu  S}(t_{2})] =\sum_{\lambda} M_{\mu \lambda} I[O_{S}(t_{1}),A_{\lambda  S}(t_{2})]. 
\end{equation}
(please see equation \eqref{eqg'}) %
As a result of the above two equations, we receive the QRT for arbitrary two-point system operators as 
\begin{equation}\label{eqh}
    \frac{d}{dt_{2}} (O(t_{1}) A_{\mu}(t_{2}))_{S} =\sum_{\lambda} M_{\mu \lambda} (O(t_{1}) A_{\lambda}(t_{2}))_{S},
\end{equation}
from which we trivially receive the standard QRT in terms of correlation functions, 
\begin{equation}\label{eqtptc}
    \frac{d}{dt_{2}} \langle O(t_{1}) A_{\mu}(t_{2})\rangle =\sum_{\lambda} M_{\mu \lambda} \langle O(t_{1}) A_{\lambda}(t_{2})\rangle,
\end{equation}
which matches with \eqref{eq8'}. Furthermore, in the expression of $I[O_{S}(t_{1}),A_{\mu  S}(t_{2})]$, it  can be shown that (see equations \eqref{eqf},\eqref{eq26'}-\eqref{eq28'}) if we swap the position of $O_{S}(t_{1})$ and $A_{\mu  S}(t_{2})$, equation \eqref{eqg} still gets respected,  i.e.,
\begin{equation}
    \frac{d}{dt_{2}} I[A_{\mu  S}(t_{2}),O_{S}(t_{1})] =\sum_{\lambda} M_{\mu \lambda} I[A_{\lambda  S}(t_{2}),O_{S}(t_{1})]
\end{equation}
and as a result we receive another form of QRT as,
\begin{equation}\label{eq1600}
    \frac{d}{dt_{2}} \langle A_{\mu}(t_{2})O(t_{1})\rangle =\sum_{\lambda} M_{\mu \lambda} \langle A_{\lambda}(t_{2})O(t_{1})\rangle.
\end{equation}
It is important to note that, if we consider the other time sequence i.e., if $t_{2}<t_{1}$ and take derivative with respect to $t_{2},$ what we receive does not obey the standard QRT. 
These results is what one also receives by working in the Schr\"odinger picture.  We next discuss the generalization of our analysis for higher-point and multi-time correlators. 

\subsection{QRT for three-point, multi-time correlation functions}
Following similar steps as before, we can define the three-point multi-time reduced operators which upto the second-order of system-bath coupling ($\lambda^2$ order) is given as \cite{karve2020heisenberg},
  \begin{widetext}
   \begin{equation} \label{eq20}
\begin{split}
(O_{1}(t_{1})O_{2}(t_{2})O_{3}(t_{3}))_{S}&=O_{1  S}(t_{1})O_{2  S}(t_{2})O_{3  S}(t_{3})+W_{1,2,3} \Big\{O_{1  S}(t_{1}) I[O_{2  S}(t_{2}),O_{3  S}(t_{3})] \Big\} \\
&+W_{1,2,3} \Big\{I[O_{1  S}(t_{1}),O_{2  S}(t_{2})] O_{3  S}(t_{3})\Big\} +W_{1,2,3} \Big\{I[O_{1  S}(t_{1}), O_{3  S}(t_{3})]O_{2  S}(t_{2})\Big\},
\end{split}
\end{equation}
\end{widetext}
 where the operator $W_{1,2,3}$ ensures that the operator product is ordered such that $O_{1S}$ comes before $O_{2S}$, and $O_{2S}$ comes before $O_{3S}$. (please see \ref{threepo} more details).  The three-point multi-time correlation functions can be computed as,
  \begin{align}\label{eqn}
     \langle O_{1}(t_{1})O_{2}(t_{2})O_{3}(t_{3})\rangle
    ={\rm Tr}_{S}[(O_{1}(t_{1})O_{2}(t_{2})O_{3}(t_{3}))_{S} \rho_{S}(0)]
  \end{align}
  
  
Now once again, to derive the QRT we first set, $O_{3}=A_{\mu}$ and assume $t_{i}<t_{3}$ with $i=1,2$. Taking  derivative of the equation \eqref{eq20} with respect to $t_{3},$ 1st term of the right hand side gives 
\begin{align}\label{eqM}
   & \frac{d}{dt_{3}} (O_{1  S}(t_{1})O_{2  S}(t_{2})A_{\mu  S}(t_{3}))\nonumber\\
   &=\sum_{\lambda} M_{\mu \lambda} (O_{1  S}(t_{1})O_{2  S}(t_{2})A_{\lambda  S}(t_{3}))
\end{align}
where we have used Eq.\eqref{eqe}.
Now once again one can show that (please see \eqref{eqN'} of appendix \ref{threereg}) for the time sequence $t_{i}<t_{3}$ with $i=1,2$ and invoking the Markovian limit, the second term of the right hand side of equation \eqref{eq20} satisfies the equation 
\begin{align} \label{eqN}
    &\frac{d}{dt_{3}}  W_{1,2,3}\{O_{1  S}(t_{1})I[O_{2  S}(t_{2}),A_{\mu  S}(t_{3})]\} \nonumber\\
    &=\sum_{\lambda} M_{\mu \lambda}  W_{1,2,3}\{O_{1  S}(t_{1})I[O_{2  S}(t_{2}),A_{\lambda S}(t_{3})]\}
\end{align}
Interestingly,  the third and the fourth term also follow identical equations like above. Finally, summing up all these contribution, we receive a multi-time QRT like form involving the reduced system operators,  
\begin{align} \label{eqOa}
   & \frac{d}{dt_{3}} (O_{1  }(t_{1})O_{2  }(t_{2})A_{\mu  }(t_{3}))_{S}\nonumber\\
   &=\sum_{\lambda} M_{\mu \lambda} (O_{1  }(t_{1})O_{2  }(t_{2})A_{\lambda  }(t_{3}))_{S}
\end{align}
from which we trivially receive the QRT for the three-point correlation functions as,
\begin{equation}\label{eq25a}
    \frac{d}{dt_{3}} \langle O_{1}(t_{1}) O_{2}(t_{2})A_{\mu}(t_{3})\rangle =\sum_{\lambda} M_{\mu \lambda} \langle O_{1}(t_{1}) O_{2}(t_{2})A_{\lambda}(t_{3})\rangle.
\end{equation} 
Interestingly if we swap the position of $O_{2}(t_{2})$ and $A_{\mu  }(t_{3})$, one can obtain a similar QRT.
The equations \eqref{eq25a} is the regression theorem for three-point correlation functions that involve three different times $t_1, t_2$, and $t_3$. This above regression type relation is of much more general form than the standard QRT derived using the Schr\"odinger picture. In fact, in special limits  $t_{1}=t_{2}$ the above QRT in \eqref{eq25a} reduces to the standard result
\begin{equation}\label{eq1066}
    \frac{d}{dt_{3}} \langle O_{1}(t_{1}) A_{\mu}(t_{3})O_{2}(t_{1})\rangle =\sum_{\lambda} M_{\mu \lambda} \langle O_{1}(t_{1}) A_{\lambda}(t_{3})O_{2}(t_{1})\rangle
\end{equation}
 Note that the QRT in our case holds irrespective of the position of $A_{\mu}(t_{3})$. However in equation \eqref{eq25a}, if we take derivative with respect to $t_{1}$ or $t_{2}$ instead of $t_{3}$ (maximum time), interestingly, we do not receive a QRT like relation.


\subsection{QRT for four-point and general N-point, multi-time correlation functions}
Following almost similar steps as before, one can work out the regression theorem for four-point functions and in fact, it is possible to generalize this analysis for $N$-point correlation functions, in the Markovian limit. Here we present the central results (see appendix \ref{fptra} for for more details). For four-point,  multi-time correlation functions we receive a QRT,
    \begin{align}\label{eqk}
    &\frac{d}{dt_{4}} \langle O_{1}(t_{1}) O_{2}(t_{2})O_{3}(t_{3})A_{\mu}(t_{4})\rangle \nonumber\\&=\sum_{\lambda} M_{\mu \lambda}  \langle O_{1}(t_{1}) O_{2}(t_{2})O_{3}(t_{3})A_{\lambda}(t_{4})\rangle.
   \end{align}
  Let us emphasize that \eqref{eqk} holds as long as $t_4> t_i$ with $i=1,2,3.$  Also note that \eqref{eqk} holds irrespective of the position of $A_{\mu}(t_{4})$ as was observed for the QRT for three-point functions. 
 This entire analysis can be generalised to $N$-point multi-time correlation functions as, 
 \begin{align}\label{eqnptr}
    &\frac{d}{dt_{N}} \langle O_{1}(t_{1}) O_{2}(t_{2})......O_{N-1}(t_{N-1})A_{\mu}(t_{N})\rangle\nonumber\\& =\sum_{\lambda} M_{\mu \lambda}  \langle O_{1}(t_{1}) O_{2}(t_{2})......O_{N-1}(t_{N-1})A_{\lambda}(t_{N})\rangle
\end{align}
 where, $t_{i}<t_{N}$ and  $i=1,2,.....N-1$. Once again, the operator $A_{\mu}(t_{N})$ can take any place and if we take derivative with respect to $t_{k}$ instead of the highest time $t_{N}$, then we do not receive the regression type formula.
 %

\section{QRT for Out-of-time-ordered correlators}\label{OTOCs}
As an application of the developed formalism, we now extend our analysis to compute the out of time ordered correlator (OTOC)\cite{otoc00} which is an excellent measure of many body localization, quantum chaos  and scrambling \cite{otoc01,otoc02,otoc03,otoc04,otoc05,otoc06,otoc07}. OTOC has received significant attention  in recent times with its applicability ranging from quantum information theory to condensed matter physics to quantum gravity. Very recently, OTOC has found its application in the context of  open quantum systems as the coupling of the system with a dissipative or dephasing bath naturally leads to information scrambling \cite{otoc3,otoc2}. Motivated by this, in this section we derive QRT like formula for OTOC correlators. Details of the derivation are provided in appendix \ref{otoc}. Here we present the main result. Let us first define four-point reduced operator of the form
\begin{equation}
(O_{1}(t_{1})A_{\mu}(t_{2})O_{3}(t_{1})A_{\nu}(t_{2}))_{S}    
\end{equation}
where the system reduced operator $A_{\mu S}(t_{2})$ satisfies Eq.~\eqref{eqe}. 
One can then receive the following regression type formula for the OTOC,
 \begin{align} \label{eqotocb}
  &\frac{d}{dt_{2}} \Big \langle O_{1}(t_{1})A_{\mu}(t_{2})O_{3}(t_{1})A_{\nu}(t_{2})\Big\rangle\nonumber\\
  &=\sum_{\lambda} M_{\mu \lambda} \Big \langle O_{1}(t_{1})A_{\lambda}(t_{2})O_{3}(t_{1})A_{\nu}(t_{2})\Big\rangle\nonumber\\
  &+\sum_{\lambda'} M_{\nu \lambda'}\Big\langle O_{1}(t_{1})A_{\mu}(t_{2})O_{3}(t_{1})A_{\lambda'}(t_{2})\Big\rangle \nonumber\\
  &+ \Big \langle W_{1,2,3,4} \Big\{O_{1  S}(t_{1})   F[A_{\mu  S}(t_{2}),A_{\nu  S}(t_{2})]O_{3  S}(t_{1})\Big\}\Big \rangle
\end{align}
where we assume $t_{2}>t_{1}$. The above equation is almost identical to the regression theorem except for the last term. 
The explicit form of the operator $ F[A_{\mu  S}(t_{2}),A_{\nu  S}(t_{2})]$ is given in the appendix (Eq.~ \eqref{eq143}). Once again, it is important to realize that this particular structure appears as a result of taking derivative with respect to $t_2$.   If we take derivative with respect to $t_{1}$ instead of $t_{2}$, such regression type form does not appear. Eq.\eqref{eqnptr} and Eq.\eqref{eqotocb} are the central results of this paper. 

Interestingly, in Ref.~\cite{otoc1}, a regression type theorem for OTOC was recently derived using a different approach. The central result obtained there is very similar to what we have derived in Eq.\eqref{eqotocb}.

%
%

\section{Example : Dissipative Spin-Boson Model}
 \label{exds}
 In this section we illustrate the above derived results for the paradigmatic dissipative spin-boson model by calculating various correlation functions in the Heisenberg picture. In particular,  we verify the QRTs for two, three and four-point correlators.  In order to perform this calculations, we first workout the master equation for reduced one point operator and then proceed to calculate the multi-time and multi-point correlation functions. The details of the derivations are provided in appendix \ref{tpds0}.
 
The Hamiltonian for a dissipative spin-1/2 system, coupled to a bath consisting of an infinite collection of harmonic oscillators with different normal mode frequencies, can be written as 
\begin{equation} \label{eq31}
\begin{split}
H & = H_{S}+H_{R}+ H_{S  R} \\
 & = \frac{\omega_{0}}{2}\sigma_{z}+\sum_{k} \omega_{k} b_{k}^\dagger b_{k}+ \sum_{k} \alpha_{k}(b_{k}^\dagger\sigma_{-}+ b_{k}\sigma_{+})\\
\end{split}
\end{equation}
where $\omega_0$ is the frequency of the qubit. The bath is charcterized by the eigen-mode frequency $\omega_k$ referring to the $k-$th oscillator with $b_{k} (b^\dagger_{k}) $ is the corresponding annihilation (creation) operator. The last term in the above Hamiltonian represents the standard dissipative coupling  term between the spin 1/2 system and the harmonic bath with $\alpha_k$ being the coupling strength between the $k-$th mode and the qubit. In what follows, for simplicity, we work in the zero-temperature limit ($T=0$).

We first derive the master equation correct upto the second-order of the system-bath coupling and further make secular approximation \cite{breuer}. One can show that the reduced one point operator obeys the following equation at zero temperature $T=0$ 
\begin{widetext}
\begin{equation} \label{eql}
\begin{split}
\frac{d}{dt}{O_{S}}(t) & =i\frac{\omega^{'}_{0}}{2} [\sigma_{z}, O_{S}(t)]+\frac{\gamma}{2}[2\sigma_{+} O_{S}(t)\sigma_{-}-\sigma_{+} \sigma_{-}O_{S}(t)-O_{S}(t)\sigma_{+}\sigma_{-}] 
\end{split}
\end{equation}
\end{widetext}
Where $\omega_{0}^{'} = \omega_0 + \Delta$ is the renormalized frequency of the qubit due to the coupling with the environment with 
\begin{equation}
\Delta={\rm P.V.}\int_{0}^{\infty}\frac{g(\omega')|\alpha(\omega')|^2d\omega'}{\omega_{0}-\omega'}.
\end{equation}
Here $g(\omega)$ represents the density of states of the bath oscillators and $\gamma= 2\pi g(\omega_{0})|\alpha(\omega_{0})|^2$ represents the net decay rate. 
Using the above master equation for the operator, it is easy to show that,
\textcolor{black}
{\begin{equation}
    \frac{d}{dt}\begin{pmatrix} \sigma_{x S}\\ \sigma_{y S} \\ \sigma_{z S} \end{pmatrix}= \begin{pmatrix}  -\frac{\gamma}{2} & -\omega'_{0} & 0 \\ \omega'_{0}& -\frac{\gamma}{2} &0 \\ 0 & 0&-\gamma \end{pmatrix}\begin{pmatrix} \sigma_{x S}\\ \sigma_{y S} \\ \sigma_{z S} \end{pmatrix} -\gamma \begin{pmatrix}  0_{2\times 2} \\ 0_{2\times 2} \\ I_{2\times 2}\end{pmatrix}
\end{equation}}
where $0_{2\times 2}$ and $I_{2\times 2}$ are the $2 \times 2$ null matrix and the identity matrix, respectively.  It is therefore easy to see that for this model there exists two different set of closed operators that follow equation \eqref{eqe}.  These are given as, 
\textcolor{black}
 { \begin{align}\label{eq33}
    A_{\mu}&= \{  \sigma_{z},I_{2\times 2}\} , ~~~ M_{\mu\lambda} = \begin{bmatrix} 
- \gamma & -\gamma \\ 0 &  0
	\end{bmatrix}\nonumber\\
	 A_{\nu}&= \{  \sigma_{x},\sigma_{y}\}, ~~~ M'_{\nu\lambda'} = 
	 \begin{bmatrix} 
 -\frac{\gamma}{2} & -\omega'_{0} \\ \omega'_{0} &  -\frac{\gamma}{2}
	\end{bmatrix}
	 \end{align}}
With this in hand, we are now ready to asses the validity of QRT results that we derived in the previous section.




%
%

 \subsection{Two-point correlation function}
Let us first verify the regression theorem for two-point function \eqref{eqh} for this model. To do that let us first set, $O_{1}=\sigma_{x}$, $A_{\mu}= \{  \sigma_{x},\sigma_{y}\}$  in equation \eqref{eqh} and assume $t_{1}<t_{2}$. Following the calculation in the Heisenberg picture for the two-point reduced operator, we receive 
\begin{widetext}
\textcolor{black}
{\begin{equation}\label{eqaaa}
    \Big(\sigma_{x}(t_{1})\sigma_{x}(t_{2})\Big)_{S}= \begin{pmatrix}  (1-\frac{\gamma}{2}(t_{2}-t_{1}))e^{-i\omega'_{0}(t_{2}-t_{1})}+2i\gamma t_{1} \sin{\omega'_{0}(t_{2}-t_{1})} & 0  \\ 0 & (1-\frac{\gamma}{2}(t_{2}-t_{1}))e^{i\omega'_{0}(t_{2}-t_{1})} \end{pmatrix},
\end{equation}
\begin{equation}\label{eqaaaa}
    \Big(\sigma_{x}(t_{1})\sigma_{y}(t_{2})\Big)_{S}= \begin{pmatrix}  i(1-\frac{\gamma}{2}(t_{2}-t_{1}))e^{-i\omega'_{0}(t_{2}-t_{1})}-2i\gamma t_{1} \cos{\omega'_{0}(t_{2}-t_{1})} & 0  \\ 0 & -i(1-\frac{\gamma}{2}(t_{2}-t_{1}))e^{i\omega'_{0}(t_{2}-t_{1})} \end{pmatrix},
\end{equation}}
and as a result the left hand side of the equation \eqref{eqh} gives 
\textcolor{black}
{\begin{equation} \label{eq13211}
\begin{split}
&\frac{d}{dt_{2}} \Big(\sigma_{x}(t_{1}) \sigma_{x}(t_{2})\Big)_{S}=-\frac{\gamma}{2}\begin{pmatrix}  e^{-i\omega'_{0}(t_{2}-t_{1})} & 0  \\ 0 & e^{i\omega'_{0}(t_{2}-t_{1})} \end{pmatrix}\\
&-\omega'_{0}\begin{pmatrix}  i(1-\frac{\gamma}{2}(t_{2}-t_{1}))e^{-i\omega'_{0}(t_{2}-t_{1})}-2i\gamma t_{1} \cos{\omega'_{0}(t_{2}-t_{1})} & 0  \\ 0 & -i(1-\frac{\gamma}{2}(t_{2}-t_{1}))e^{i\omega'_{0}(t_{2}-t_{1})} \end{pmatrix}
\end{split}
\end{equation}}

Similarly, the right hand side of the equation \eqref{eqh} can be computed and we receive,
\textcolor{black}
{\begin{align}\label{eqA6611}
   & -\frac{\gamma}{2} \Big(\sigma_{x}(t_{1}) \sigma_{x}(t_{2})\Big)_{S}-\omega'_{0}\Big(\sigma_{x}(t_{1})\sigma_{y}(t_{2})\Big)_{S} \nonumber\\
   &=-\frac{\gamma}{2}\begin{pmatrix}  e^{-i\omega'_{0}(t_{2}-t_{1})} & 0  \\ 0 & e^{i\omega'_{0}(t_{2}-t_{1})} \end{pmatrix}\\
&-\omega'_{0}\begin{pmatrix}  i(1-\frac{\gamma}{2}(t_{2}-t_{1}))e^{-i\omega'_{0}(t_{2}-t_{1})}-2i\gamma t_{1} \cos{\omega'_{0}(t_{2}-t_{1})} & 0  \\ 0 & -i(1-\frac{\gamma}{2}(t_{2}-t_{1}))e^{i\omega'_{0}(t_{2}-t_{1})} \end{pmatrix}
\end{align}}
\end{widetext}
These above two equations therefore shows the validity of QRT at the level of two-point reduced operators. As an immediate consequence, we conclude that  the QRT follows for two-point correlation functions Eq.\eqref{eqtptc} for arbitrary initial density matrix for the system.



 \subsection{Three-point, multi-time correlation function}


We  next move to three-point correlation function as given in Eq.~\eqref{eqOa}. We first set, $O_{1}=O_{2}=\sigma_{x}$, $A_{\mu}= \{  \sigma_{x},\sigma_y\}$ and assume $t_{1}<t_{2}<t_{3}$. The left hand side of the equation \eqref{eqOa} can be computed following the Heisenberg picture (please see the details of the calculation in appendix \ref{tpds0}),
 \begin{widetext}
 \textcolor{black}
{\begin{equation}
    (\sigma_{x}(t_{1}) \sigma_{x}(t_{2}) \sigma_{x}(t_{3}))_{S}=\\
    \begin{pmatrix} 0  & \big(1-\frac{1}{2}\gamma (t_{1}+t_{3}-t_{2})\big)e^{i\omega'_{0}(t_{1}+t_{3}-t_{2})}\\ \big(1-\frac{1}{2}\gamma (-t_{1}+t_{2}+t_{3})\big)e^{-i\omega'_{0}(t_{1}+t_{3}-t_{2})}\\
   + \gamma (t_{2}-t_{1})e^{i\omega'_{0}(-t_{1}-t_{2}+t_{3})} &    0 \end{pmatrix}
\end{equation}
and
\begin{align}
    (\sigma_{x}(t_{1}) \sigma_{x}(t_{2}) \sigma_{y}(t_{3}))_{S}=\begin{pmatrix} 0  & -i\big(1-\frac{1}{2}\gamma (t_{1}+t_{3}-t_{2})\big)e^{i\omega'_{0}(t_{1}+t_{3}-t_{2})}\\ i\big(1-\frac{1}{2}\gamma (-t_{1}+t_{2}+t_{3})\big)e^{-i\omega'_{0}(t_{1}+t_{3}-t_{2})}\\
    -i\gamma (t_{2}-t_{1})e^{i\omega'_{0}(-t_{1}-t_{2}+t_{3})} &    0 \end{pmatrix}
\end{align}
}
Now taking derivative of the first equation with respect to the maximum time $t_{3}$, we get,
\textcolor{black}
{\begin{align}
   & \frac{d}{dt_{3}}(\sigma_{x}(t_{1}) \sigma_{x}(t_{2}) \sigma_{x}(t_{3}))_{S}=-\frac{1}{2} \gamma\begin{pmatrix} 0  & e^{i\omega'_{0}(t_{1}+t_{3}-t_{2})}\\ e^{-i\omega'_{0}(t_{1}+t_{3}-t_{2})} &    0 \end{pmatrix}\nonumber\\
   &-\omega'_{0}  \begin{pmatrix} 0  & -i\big(1-\frac{1}{2}\gamma (t_{1}+t_{3}-t_{2})\big)e^{i\omega'_{0}(t_{1}+t_{3}-t_{2})}\\ i\big(1-\frac{1}{2}\gamma (-t_{1}+t_{2}+t_{3})\big)e^{-i\omega'_{0}(t_{1}+t_{3}-t_{2})}\\
   -i\gamma (t_{2}-t_{1})e^{i\omega'_{0}(-t_{1}-t_{2}+t_{3})} &    0 \end{pmatrix} 
     \label{three-LHS}
\end{align}
Now, one can calculate the RHS of \eqref{eqOa} and obtain, 
\begin{align}
   &  -\frac{1}{2} \gamma \,  (\sigma_{x}(t_{1}) \sigma_{x}(t_{2}) \sigma_{x}(t_{3}))_{S}-\omega'_{0} \, (\sigma_{x}(t_{1}) \sigma_{x}(t_{2}) \sigma_{y}(t_{3}))_{S}\nonumber\\
   &= -\frac{1}{2} \gamma\begin{pmatrix} 0  & e^{i\omega'_{0}(t_{1}+t_{3}-t_{2})}\\ e^{-i\omega'_{0}(t_{1}+t_{3}-t_{2})} &    0 \end{pmatrix}\nonumber\\
   &-\omega'_{0}  \begin{pmatrix} 0  & -i\big(1-\frac{1}{2}\gamma (t_{1}+t_{3}-t_{2})\big)e^{i\omega'_{0}(t_{1}+t_{3}-t_{2})}\\ i\big(1-\frac{1}{2}\gamma (-t_{1}+t_{2}+t_{3})\big)e^{-i\omega'_{0}(t_{1}+t_{3}-t_{2})}\\
   -i\gamma (t_{2}-t_{1})e^{i\omega'_{0}(-t_{1}-t_{2}+t_{3})} &    0 \end{pmatrix}
      \label{three-RHS}
\end{align}}
\end{widetext}
Equations  \eqref{three-LHS} and \eqref{three-RHS} gives the identical result, this verifies the regression theorem \eqref{eqOa}. Note that, we choose a particular order of time but note that we can show the equation \eqref{eqOa} holds as long as $t_{i}<t_{3}$ with $i=1,2$. There are no constrains on the order of $(t_{1},t_{2})$.

 \subsection{Four-point, multi-time correlation function}
 Now we want to verify the regression theorem for four-point function \eqref{eqk} in this example. To do that let us first set, $O_{1}=O_{2}=O_{3}=\sigma_{x}$, $A_{\mu}= \{  \sigma_{x},\sigma_y\}$  in equation \eqref{eqk} and assume $t_{1}<t_{2}<t_{3}<t_{4}$. We then compute the following reduced operators which  turns out to be diagonal. More explicitly,  
 \begin{widetext}
 \begin{equation}\label{eq4ptr}
\begin{split}
   (\sigma_{x}(t_{1}) \sigma_{x}(t_{2}) \sigma_{x}(t_{3})\sigma_{x}(t_{4}))_{S}&= \begin{pmatrix}  \big(1-\frac{1}{2}\gamma (t_{1}+t_{3}-t_{2}-t_{4})\big)e^{i\omega'_{0}(t_{1}+t_{3}-t_{2}-t_{4})}\\
  +\gamma t_{1}e^{i\omega'_{0}(-t_{1}+t_{2}-t_{3}+t_{4})}\\
  +\gamma (t_{3}-t_{2})e^{i\omega'_{0}(t_{1}-t_{2}-t_{3}+t_{4})} & 0\\ 0&
  \big(1-\frac{1}{2}\gamma (t_{1}+t_{2}+t_{3}+t_{4})\big)e^{i\omega'_{0}(-t_{1}+t_{2}-t_{3}+t_{4})}\\
  &+\gamma t_{1}e^{i\omega'_{0}(-t_{1}+t_{2}-t_{3}+t_{4})}\\
 & +\gamma t_{3}e^{i\omega'_{0}(-t_{1}+t_{2}-t_{3}+t_{4})} \end{pmatrix}
   \end{split}
\end{equation}
and,
\begin{equation}\label{eq4ptr1}
\begin{split}
   (\sigma_{x}(t_{1}) \sigma_{x}(t_{2}) \sigma_{x}(t_{3})\sigma_{y}(t_{4}))_{S}&= \begin{pmatrix}  i\big(1-\frac{1}{2}\gamma (t_{1}+t_{3}-t_{2}-t_{4})\big)e^{i\omega'_{0}(t_{1}+t_{3}-t_{2}-t_{4})}\\
  -i\gamma t_{1}e^{i\omega'_{0}(-t_{1}+t_{2}-t_{3}+t_{4})}\\
  -i\gamma (t_{3}-t_{2})e^{i\omega'_{0}(t_{1}-t_{2}-t_{3}+t_{4})} & 0\\ 0&
 -i \big(1-\frac{1}{2}\gamma (t_{1}+t_{2}+t_{3}+t_{4})\big)e^{i\omega'_{0}(-t_{1}+t_{2}-t_{3}+t_{4})}\\
  &-i\gamma t_{1}e^{i\omega'_{0}(-t_{1}+t_{2}-t_{3}+t_{4})}\\
 & -i+\gamma t_{3}e^{i\omega'_{0}(-t_{1}+t_{2}-t_{3}+t_{4})} \end{pmatrix}
   \end{split}
\end{equation}
Taking derivative of the equation \eqref{eq4ptr} with respect to $t_{4}$, we receive the QRT as, 
\begin{equation}\label{eq1}
\begin{split}
   \frac{d}{dt_{4}}(\sigma_{x}(t_{1}) \sigma_{x}(t_{2}) \sigma_{x}(t_{3})\sigma_{x}(t_{4}))_{S}&=-\frac{1}{2} \gamma\,  \big(\sigma_{x}(t_{1}) \sigma_{x}(t_{2} \big) \sigma_{x}(t_{3}))\sigma_{x}(t_{4}))_{S}\\
   &-\omega'_{0} \big(\sigma_{x}(t_{1}) \sigma_{x}(t_{2}) \sigma_{x}(t_{3})\sigma_{y}(t_{4})\big)_{S}\\
   \end{split}
   \end{equation}
  \end{widetext}



This immediately verifies equation \eqref{eqk}. Note that, we choose here a particular order of time but note that one can show the validity of the QRT as long as $t_{i}<t_{4}$ with $i=1,2,3$. There are no constrains on the order of $(t_{1},t_{2},t_{3})$.

\subsection{Verification of OTOC}
We next provide one example to asses the validity of our expression for OTOC. For the spin-boson model, it is easy to compute the following four-point reduced operator and one receives,
\textcolor{black}
{\begin{equation}
   \frac{d}{dt_{2}}(\sigma_{x}(t_{1}) \sigma_{z}(t_{2}) \sigma_{x}(t_{1})\sigma_{z}(t_{2}))_{S}=2 \gamma \begin{pmatrix} 1 & 0\\  0 &   1 \end{pmatrix}.
\end{equation}}
It is easy to check that the corresponding RHS of the OTOC also gives the same result. In a similar way, OTOC can be checked for 
\textcolor{black}
{\begin{equation}
   \frac{d}{dt_{2}}(\sigma_{z}(t_{1}) \sigma_{z}(t_{2}) \sigma_{z}(t_{1})\sigma_{z}(t_{2}))_{S}=-8 \gamma \begin{pmatrix} 1 & 0\\  0 &   0 \end{pmatrix}.
\end{equation}}

\section{Generalization of QRT for non-Markovian case}
\label{nmv}
The results for QRT presented in the previous sections can be extended for the non-Markovian case. For simplicity, We here focus on systems with bosonic bath and linear system-bath interaction but one can generalize this study for a more generic type of system-bath interaction as well. We  derive a Lindblad type equation upto order $\lambda^2$ for this setup which takes into account the non-Markovian evolution.  We then derive the correction to the QRT for the non-Markovian case by focusing only on the two-point correlation functions \cite{PhysRevA.75.052108}. Extension to higher point multi-time correlators can be similarly obtained even for the non-Markovian case. 
\subsection{Lindblad type non-Markovian equation for one-point reduced operator}
Let the Hamiltonian of the composite system+ bath is,
\begin{equation} \label{eq1nm}
\begin{split}
H & = H_{S}+H_{R}+ H_{S  R} \\
 & = H_{S}+\sum \omega_{k} b_{k}^\dagger b_{k}+ \sum g_{k}( L b_{k}^\dagger+L^\dagger   b_{k}).
\end{split}
\end{equation}
Here, the system is coupled with the bath through a generic system operator $L$. Let us assume that the initial density operator of the total system can be written as, $\rho_{SR}(0)=\rho_{S} \otimes \rho_{R}.$ Now following the master equation in the Heisenberg picture, (please see equation \eqref{eq74}), one can show that the reduced density operator for the system obeys the following non-Markovian master equation at zero temperature ($T=0$) and is correct up to the second order of system-bath coupling,
\begin{equation} \label{eqsa}
\begin{split}
\frac{d}{dt}{O_{S}}(t) & = i [H_{S}, O_{S}(t)]+ \int^{t}_0 d\tau \, \alpha(\tau) \Big[L^\dagger(0), O_{S}(t)\Big] \Tilde{L}(-\tau) \\
 & +  \int^{t}_0 d\tau  \, \alpha^*(\tau)\Tilde{L}^\dagger(-\tau) \Big[O_{S}(t),{L}(0)\Big],
\end{split}
\end{equation}
where,  $ \Tilde{L}(t)=U_{0}(t) LU^\dagger_{0}(t)$ is the coupled system operator in the interaction picture with $U_0(t)$ represents the free evolution due to the Hamiltonian $H_S$. $ \alpha(\tau)$ denotes the bath correlation function which is given as,
\begin{equation}
   \alpha(\tau)=\sum_{k}  |g_k|^2 \, {\rm Tr}_{R} \Big[\Tilde{b}_{  k}(0)\Tilde{b}^\dagger_{k}(-\tau)\rho_{R}\Big] =\sum_{k}  |g_k|^2e^{-i\omega_{k}\tau}.
\end{equation}
\subsection{Extension of QRT to non-Markovian case}
Having obtained the non-Markovian master equation in the schr\"odinger picture, we now extend the QRT for the non-Markovian dynamics. To achieve that let us first  assume that there exists a complete set of system operators $A_{\mu  S}(t)$ such that 
 \begin{equation}\label{eqva}
     \frac{d}{dt} A_{\mu  S}(t)=\sum_{\lambda} M_{\mu  \lambda}(t) \, A_{\lambda  S}(t)
  \end{equation}
 Note the crucial explicit time dependence in $M_{\mu  \lambda}(t)$ for the non-Markovian case which was independent of time for the Markovian dynamics.
 Using the above results, it is easy to derive QRT for the non-Markovian case and is given by (see appendix \ref{nmvqrt} for derivation)
\begin{equation} \label{eq1nva}
\begin{split}
 &\frac{d}{dt_{1}} \left (A_{\mu}(t_{1})O(t_{2})\right)_{S} =\sum_{\lambda} M_{\mu \lambda}(t_{1}) \left(A_{\lambda}(t_{1})O(t_{2})\right)_{S}\\
   &-  \int_{0}^{t_{2}}d\tau_{2}\; \alpha(\tau_{2}\!-\!t_{1})\,A_{\mu  S}(t_{1})
 \Tilde{ L}^\dagger(-t_{1}) O_{ S}(t_{2})
   \Tilde{L}(-\tau_{2})\\
   &- \int_{0}^{t_{2}}d\tau_{2}\; \alpha(\tau_{2}\!-\!t_{1}) \, \Tilde{ L}^\dagger(-t_{1})A_{\mu  S}(t_{1})
  \Tilde{L}(-\tau_{2})O_{  S}(t_{2})
  \\
   &+  \int_{0}^{t_{2}}d\tau_{2}\; \alpha(\tau_{2}\!-\!t_{1})\, \Tilde{ L}^\dagger(-t_{1})A_{\mu  S}(t_{1})
 O_{  S}(t_{2})
   \Tilde{L}(-\tau_{2})\\
   &+  \int_{0}^{t_{2}}d\tau_{2}\; \alpha(\tau_{2}\!-\!t_{1})\,A_{\mu  S}(t_{1})
 \Tilde{ L}^\dagger(-t_{1}) \Tilde{L}(-\tau_{2}) O_{  S}(t_{2})
  \end{split}
\end{equation}
where we have assumed $t_{1}>t_{2}$ and that the operator $A_{\mu}$ satiesfies Eq.~\eqref{eqva}.
The above equation is the extension of QRT to the non-Markovian dynamics. In the Markovian limit Eq.~\eqref{eq1nva} correctly reproduce the standard QRT, as given in Eq.~\eqref{eq14}. Note that, In the Markovian limit, the bath correlation function $\alpha(\tau_{2}-t_{1})$ is  non-vanishing only if $\tau_{2}$ lies in the range from $t_{1}-\tau_{B}$ to $t_{1}+\tau_{B}$ where $\tau_B$ corresponds to a characteristic time scale of the  bath dynamics. However, the upper limit of the integration in equation (\ref{eq1nva}) is $t_{2}$  which is always less than $t_{1}$. This implies the bath correlation function, $\alpha(\tau_{2}-t_{1}) \to 0$ in the given range of $\tau_{2}$.  Eq.~(\ref{eq1nva}) thus reduces to the QRT for two-point function with the last four terms disapper in the Markovian limit.

\section{Discussion}
While defining the multi-time correlation function, it is natural to work in the Heisenberg picture. Even though, multi-time correlation function is of great physical importance, Heisenberg picture has not got that much attention in the context of quantum open system. In this paper we make use of recently developed Heisenberg Picture technique \cite{karve2020heisenberg} to calculate correlation functions in the Markovian limit and derive quantum regression theorem. In particular, we generalize the regression theorem for multi-time correlation functions with general time arguments. What we observe is that that the form of regression theorem remains the same for two or multi-time correlation functions as long as some mild restriction on the time arrangements is met.  We also derived regression theorem for OTOC in the Markovian limit and find that regression theorem gets modification from the known two-time regression theorem. We further extend our study to the non-Markovian dynamics. However in this case the QRT receives a complicated correction term with two-point correlation function requiring information about four-point function. 

As a possible future direction, one of the interesting problem would be to use the Heisenberg picture \cite{karve2020heisenberg} and to go beyond the standard second-order perturbation scheme. For example, one can consider exactly solvable systems such as the spin-boson dephasing model, or the well-known Caldeira-Leggett model and  investigate the possibility to sum up the  perturbation series exactly in the Heisenberg picture. 

\textbf{Acknowledgements:} 
Work of SJ is supported by Ramanujan Fellowship. Work of SK is supported by CSIR fellowship with Grant Number 09/0936(11643)/2021-EMR-I.   BKA acknowledges the MATRICS grant
MTR/2020/000472 from SERB, Government of India. BKA also thanks the Shastri Indo-Canadian
Institute for providing financial support for this research
work in the form of a Shastri Institutional Collaborative
Research Grant (SICRG). The authors would also like to thank the people of India for their steady support in basic research.

 \bibliographystyle{apsrev4-1}
 \bibliography{abc.bib}

\appendix
\newpage
\section{Review of Heisenberg picture dynamics}\label{Heisencor}
 In this Appendix for completeness, we briefly review the Heisenberg picture results as obtained by \cite{karve2020heisenberg}. We follow the notation and convention introduced in this reference . Let the Hamiltonian of the total system is $H  = H_{S}+H_{R}+\lambda \, H_{S  R}.$ where we introduce an extra parameter $\lambda$ to keep track of the order of perturbation in terms of system-bath coupling. 
Let us assume that the interaction between system and bath is turned on  at $t=0.$ Before turning the interaction, system and bath were decoupled and their total density matrix can be written as, $\rho_{SR}=\rho_{S} \otimes \rho_{R}.$ In the  Heisenberg picture the density matrix is time independent. The expectation value of any operator can be written as,
 \begin{equation}
      \langle O(t)\rangle={\rm Tr}_{S}[{\rm Tr}_{R}[O(t)\rho_{R}]\rho_{S}]={\rm Tr}_{S}[O_{S}(t)\rho_{S}]
  \end{equation}
 where the reduced one point operator define as, $O_{S}(t)= Tr_{R}[O(t)\rho_{R}]$. Similarly, we can define $N$-point reduced operator as
 \begin{eqnarray}\label{redoa}
&& (O_{1}(t_{1})O_{2}(t_{2})....O_{N}(t_{N}))_{S} = \nonumber \\
&& \quad \quad {\rm Tr}_{R}[O_{1}(t_{1})O_{2}(t_{2})....O_{N}(t_{N})\rho_{R}].
 \end{eqnarray}
This definition has a property that  $(O_{1}(t_{1})O_{2}(t_{2}))_{S}\neq O_{1  S}(t_{1})O_{2  S}(t_{2})$
 However, we can express the reduced $N$-point operator in terms of one-point reduced operators using what are called image operators \cite{karve2020heisenberg}. The image operator of any operator $O(t)$ is defined as,
 \begin{equation}\label{eq3}
     O_{\alpha \beta}(t)=T^\dagger_{\alpha}O(t)T_{\beta}
 \end{equation}
 where, $T_{\alpha}=\ket{i\alpha}\bra{i}$ where $\{\ket{i}\}$ are orthonormal basis of $H_{S}$ and  $\{\ket{\alpha}\}$ are  orthonormal basis of $H_{R}$. It is easy to see that
 \begin{equation}T_{\alpha}T^\dagger_{\alpha}=1.\label{ttd}
 \end{equation}One can show that the $N$-point image operators can be written in terms of one-point image operators as
 \begin{align}\label{im1pt}
 &(O_{1}(t_{1})O_{2}(t_{2})....O_{N}(t_{N}))_{\alpha\beta}\nonumber\\
 &= O_{1  \alpha\gamma_{1}}(t_{1})O_{2  \gamma_{1}\gamma_{2}}(t_{2})....O_{N  \gamma_{N-1}\beta}(t_{N}).
 \end{align}
The $N$-point reduced operators defined in \eqref{redoa} can also be expressed interms of one point image operators by inserting \eqref{ttd} and using \eqref{im1pt}  as follows\footnote{We can explicitely express the one-point image operator in terms of one point reduced operator\cite{karve2020heisenberg}.}
 \begin{align}\label{eqnptro}
 & (O_{1}(t_{1})O_{2}(t_{2})....O_{N}(t_{N}))_{S} \nonumber\\
 &=O_{1  \alpha\gamma_{1}}(t_{1})O_{2  \gamma_{1}\gamma_{2}}(t_{2})....O_{N  \gamma_{N-1}\beta}(t_{N})\rho_{R  \beta\alpha}
  \end{align}
 
 We next consider a general form of the interaction Hamiltonian between system and bath and write
 \begin{equation}
    H_{S  R}=\sum_{i} S^i\otimes R^i
\end{equation}
where, $S^i$ is a Hermitian operator acting on the system’s Hilbert space, and $ R^i$
is a hermitian operator in the bath’s Hilbert space. The corresponding image operators of $H_{S  R}$ are (using equation \eqref{eq3}):
\begin{equation}
    H_{S  R  \alpha\gamma}=\sum_{i}S^iR^i_{\alpha\gamma},
\end{equation}
Here, $R^i_{\alpha\gamma}$ is the $\alpha$, $\gamma$ th element of  $R^i$ \big($\ket{\alpha}$, $\ket{\gamma}$ are the eigenstates of bath Hamiltonian $ H_{R}$\big). We define the interaction picture operators $ \Tilde{H}_{  S  R  \alpha\gamma}(t)$ as,
\begin{equation}
    \tilde{H}_{ S R \alpha\gamma}(t)=\sum_{i}\Tilde{S^i}(t)\Tilde{R^i}_{  \alpha\gamma}(t),
\end{equation}
where, $ \Tilde{S^i}(t)=U_{0}(t) S^iU^\dagger_{0}(t)=\sum_{\omega} S^i_{\omega} e^{i \omega t}$ and $\Tilde{R^i}_{  \alpha\gamma}(t)=R^i_{\alpha\gamma}e^{-i(E_{\alpha}-E_{\gamma})t}$;  here $
 U_{0}(t)=e^{-iH_{S}t}$ and  $E_{\alpha}$, $E_{\gamma}$ are the eigenvalues of bath Hamiltonian $ H_{R}$.  Then the exact equation (written to all orders in $\lambda$) that satisfies the reduced one point system operator $O_{S}(t)$  is\cite{karve2020heisenberg}, 
 \begin{widetext}
\begin{align}\label{eq74}
   \frac{d}{dt}O_{S}(t)=i[H_{S},O_{S}(t)]+\sum_{n=1}^\infty \sum_{k=0}^\infty \sum_{n_{1},n_{2},..n_{k}=1}^\infty (-1)^k \lambda^{n+n_{1}+...+n_{k}} D_{t}P_{s}^nP_{s}^{n_{1}}....P_{s}^{n_{k}} O_{S}(t) 
\end{align}    
\end{widetext}
where the super-operator $D_{t}P_{s}^n$ is defined as 
\footnote{Throughout the paper we are going to use the Einstein summation convention that repeated indices are summed.}
\begin{widetext}
   \begin{equation} \label{eq111}
\begin{split}
D_{t}P_{s}^n A(t) & = \sum_{r=0}^n i^{n-2r} U_{0}^\dagger(t) \frac{d}{dt}[U_{0}(t) K_{\gamma  \alpha}^{(n-r) \dagger}(t) U_{0}^\dagger(t)]U_{0}(t) A(t) K_{\gamma  \beta}^r(t) \rho_{B  \beta  \alpha}\\
 &+\sum_{r=0}^n i^{n-2r}  K_{\gamma  \alpha}^{(n-r)\dagger}(t) A(t)U_{0}^\dagger(t) \frac{d}{dt}[U_{0}(t) K_{\gamma  \beta}^r(t) U_{0}^\dagger(t)]U_{0}(t)  \rho_{B  \beta  \alpha}
\end{split}
\end{equation}
\end{widetext}
 here,
\begin{align}\label{kdef}
  K_{\alpha \beta}^r(t)&=e^{i(E_{\alpha}-E_{\beta})t}U_{0}^\dagger(t) \tilde{K}_{  \alpha  \beta}^r(t) U_{0}(t), \nonumber\\
   \tilde{K }_{ \alpha  \beta}^n(t)&=\int_{0}^t dt_{1}....\int_{0}^{t_{n-1}} dt_{n} \tilde{H}_{  S  R  \alpha  \gamma_{1}}(t_{1})....\tilde{H }_{ S  R  \gamma_{n-1}  \beta}(t_{n}).
\end{align}
with
\begin{equation}\label{k0}
      \tilde{K }_{\alpha \beta}^0(t)=\delta_{\alpha \beta}.
\end{equation}

 \subsection{Correlation function in the Heisenberg picture}
\subsubsection{Expression for two-point reduced operators}
We now compute the  two-point reduced operator in the Heisenberg picture which can  be written as \cite{karve2020heisenberg}:
  \begin{equation}\label{eq2}
    ( O_{1}(t_{1})O_{2}(t_{2}))_{S}=O_{1  S}(t_{1})O_{2  S}(t_{2})+
 I[O_{1  S}(t_{1}),O_{2  S}(t_{2})] 
  \end{equation}
 Where $I[O_{1  S}(t_{1}),O_{2  S}(t_{2})] $ is the irreducible part\footnote{it can't be expressed simply as the multiplication of two one pint reduced operator but it's a function of one point reduced operator and it starts from $\lambda^2$ order}. This irreducible part can be expressed upto $\lambda^2$ order, following Eq.\eqref{eqnptro} as,
 \begin{widetext}
 \begin{equation} \label{eqI}
\begin{split}
   I[O_{1  S}(t_{1}),O_{2  S}(t_{2})] =
  \sum_{n_{0}^1,l_{0}^1,n_{0}^2,l_{0}^2} &\left((-i\lambda)^{n_{0}^1} K^{n_{0}^1 }_{ \gamma_{0} \alpha}(t_{1})\right)^\dagger O_{1  S}(t_{1}) \left((-i\lambda)^{l_{0}^1} K^{l_{0}^1}_{  \gamma_{0} \gamma}(t_{1})\right)  \\
  & \left((-i\lambda)^{n_{0}^2}K^{n_{0}^2}_{  \gamma'_{0} \gamma}(t_{2})\right)^\dagger O_{2  S}(t_{2})
  \left((-i\lambda)^{l_{0}^2} K^{l_{0}^2 }_{ \gamma'_{0} \beta}(t_{2})\right) \rho_{R \beta\alpha}
\end{split}
\end{equation}
\end{widetext}
such that $n_{0}^1+l_{0}^1=1$ and $n_{0}^2+l_{0}^2=1$, so these are the following four possible  combinations,
\begin{enumerate}
\item $n_{0}^1=0,l_{0}^1=1 $ and $n_{0}^2=0,l_{0}^2=1$
\item $n_{0}^1=1,l_{0}^1=0 $ and $n_{0}^2=0,l_{0}^2=1$
\item $n_{0}^1=0,l_{0}^1=1 $ and $n_{0}^2=1,l_{0}^2=0$
\item $n_{0}^1=1,l_{0}^1=0 $ and $n_{0}^2=1,l_{0}^2=0$
\end{enumerate}
We can then write $I$ as the sum over these four combinations i.e.,
\begin{equation}\label{eqf}
        I=I_{1}+I_{2}+I_{3}+I_{4}.
    \end{equation}
One can write down the expressions for each $I_i$. For example, using \eqref{kdef}, \eqref{k0},\eqref{eqI}, we obtain
\begin{widetext}
 \begin{align}\label{eqI4}
 I_{1}=-\lambda^2\sum_{\omega,\omega'}\sum_{i,j}O_{1  S}(t_{1}) S^i_{\omega} O_{2  S}(t_{2})S^j_{\omega'}\int^{t_{1}}_{  0}d\tau_{1}e^{-i\omega\tau_{1}} \int^{t_{2} }_{ 0}  d\tau_{2} e^{-i \omega'\tau_{2}} Tr_{R} [\Tilde{R}^i(-\tau_{1})\Tilde{R}^j(-\tau_{2})\rho_{R}]
\end{align}
\end{widetext}
where, $\tau_{1}=t_{1}-t'_{1}$ and  $\tau_{2}=t_{2}-t'_{2}$.

Similarly, we can show that the other contributions gives,
\begin{widetext}
\begin{equation} \label{eqH}
\begin{split}
 I_{2}=&-\lambda^2 \sum_{\omega,\omega'} \sum_{i,j} S^{i \dagger}_{  \omega}O_{1  S}(t_{1})
   O_{2  S}(t_{2})
  S^j_{\omega'}\int^{t_{1} }_{ 0}d\tau_{1}e^{i\omega\tau_{1}}\int^{t_{2}}_{  0}  d\tau_{2} e^{-i\omega'\tau_{2}}  \, {\rm Tr}_{R} [\Tilde{R}  ^{i } (-\tau_{1})\Tilde{R}^j(-\tau_{2})\rho_{R}] \\
 I_{3}=&-\lambda^2\sum_{\omega,\omega'}\sum_{i,j}O_{1  S}(t_{1}) S^i_{\omega} O_{2  S}(t_{2})S^{j \dagger}_{  \omega'}\int^{t_{1}}_{  0}d\tau_{1}\;e^{-i\omega\tau_{1}} \int^{t_{2} }_{ 0}  d\tau_{2} e^{i\omega'\tau_{2}} \, {\rm Tr}_{R} [\Tilde{R}^i (-\tau_{1})\Tilde{R}  ^{j }(-\tau_{2})\rho_{R}]\\
 I_{4}=&-\lambda^2 \sum_{\omega,\omega'} \sum_{i,j} S^{i \dagger}_{  \omega} O_{1 S}(t_{1})
  S^{j \dagger}_{  \omega'}O_{2 S}(t_{2})
   \int^{t_{1}}_{  0}d\tau_{1}e^{i\omega\tau_{1}} \int^{t_{2} }_{ 0}  d\tau_{2} ~e^{i\omega'\tau_{2}} \, {\rm Tr}_{R} [\Tilde{R}  ^{i } (-\tau_{1})\Tilde{R}  ^{j }(-\tau_{2})\rho_{R}] \\
\end{split}
\end{equation}
\end{widetext}
Note that, all these expressions are correct upto order $\lambda^2$ and is valid for general dynamics.

\subsubsection{Expression for three-point reduced operator}\label{threepo}
 As before, we can work out expressions for three point reduced operator upto $\lambda^2$ order. We receive \cite{karve2020heisenberg}
\begin{align} \label{eq22}
&(O_{1}(t_{1})O_{2}(t_{2})O_{3}(t_{3}))_{S}\nonumber\\
&=O_{1  S}(t_{1})O_{2  S}(t_{2})O_{3  S}(t_{3}) \nonumber \\
&+W_{1,2,3} \Big\{I[O_{1  S}(t_{1}),O_{2  S}(t_{2})] O_{3  S}(t_{3})\Big\}\nonumber\\
& +W_{1,2,3} \Big\{O_{1  S}(t_{1}) I[O_{2  S}(t_{2}),O_{3  S}(t_{3})] \Big\}\nonumber\\
&+W_{1,2,3} \Big\{I[O_{1  S}(t_{1}), O_{3  S}(t_{3})]O_{2  S}(t_{2})\Big\}
\end{align}
 The operator $ W_{1,2,3}$ makes sure that the operator product is ordered such that $O_{1 S}$ comes before $O_{2 S}$, and $O_{2 S}$ comes before $O_{3 S}$. Let us illustrate this by one example. Considering the last term of the above equation \eqref{eq22} we get,
  \begin{align} \label{eq85}
 &W_{1,2,3} \Big\{I[O_{1  S}(t_{1}), O_{3  S}(t_{3})]O_{2  S}(t_{2})\Big\} \nonumber\\
 &= \lambda^2\sum_{n_{0}^1,l_{0}^1,n_{0}^2,l_{0}^2}i^{n_{0}^1+n_{0}^2-l_{0}^1-l_{0}^2} \left( K^{n_{0}^1 }_{ \gamma_{0} \alpha}(t_{1})\right)^\dagger O_{1  S}(t_{1})  K^{l_{0}^1 }_{ \gamma_{0} \gamma}(t_{1}) O_{2  S}(t_{2})\nonumber \\
  & ~~~~~~~~~~~~~~~~~~~~~~~~~~~~~~~~~*\left(K^{n_{0}^2  }_{\gamma'_{0} \gamma}(t_{3})\right)^\dagger O_{3  S}(t_{3})
   K^{l_{0}^2 }_{ \gamma'_{0} \beta}(t_{3}) \rho_{R  \beta \alpha}
\end{align}
such that $n_{0}^1+l_{0}^1=1$ and $n_{0}^2+l_{0}^2=1$.
  \subsubsection{Expression for four-point reduced operators}
We can now calculate the four point function as well. The four-point reduced operator  (upto $\lambda^2$ order) is\cite{karve2020heisenberg},
\begin{widetext}
 \begin{equation} \label{eq88}
\begin{split}
&(O_{1}(t_{1})O_{2}(t_{2})O_{3}(t_{3})O_{4}(t_{4}))_{S}\\&=O_{1  S}(t_{1})O_{2  S}(t_{2})O_{3  S}(t_{3})O_{4  S}(t_{4}) +W_{1,2,3,4} \Big\{I[O_{1  S}(t_{1}),O_{2  S}(t_{2})] O_{3  S}(t_{3})O_{4  S}(t_{4})\Big\} \\
     & +W_{1,2,3,4} \Big\{O_{1  S}(t_{1}) I[O_{2  S}(t_{2}),O_{3  S}(t_{3})] O_{4  S}(t_{4})\Big\} +W_{1,2,3,4} \Big\{I[O_{1  S}(t_{1}), O_{3  S}(t_{3})]O_{2  S}(t_{2})O_{4  S}(t_{4})\Big\}\\
     &+W_{1,2,3,4} \Big\{I[O_{4  S}(t_{4}), O_{3  S}(t_{3})]O_{2  S}(t_{2})O_{1  S}(t_{1})\Big\} 
    + W_{1,2,3,4} \Big\{I[O_{1  S}(t_{1}),O_{4  S}(t_{4})] O_{3  S}(t_{3})O_{2  S}(t_{2})\Big\} \\
     &+W_{1,2,3,4} \Big\{O_{1  S}(t_{1}) I[O_{2  S}(t_{2}),O_{4  S}(t_{4})] O_{3  S}(t_{3})\Big\} 
\end{split}
\end{equation}
\end{widetext}
where functions $W$ and $I$ has the same property as is already discussed previously.
For our purpose we need to calculate the explicit form of the function $I.$ Below we give an explicit example.

\subsection{Results in the Markovian limit}
\label{markov1}

In the Markovian limit\cite{karve2020heisenberg} the bath correlation functions i.e., ${\rm Tr}_{R}[\Tilde{R}(t)\Tilde{R}(t-\tau)\rho_{R}]$ is a rapidly decaying function of $\tau$ only, using this property we can show that the general equation \eqref{eq74} reduces to the well-known master equation for reduced operator 
\cite{karve2020heisenberg}
\begin{align}\label{eq14a}
	&\frac{d}{dt}O_{S}(t)=i H_{S} O_{S}(t) \nonumber\\
	&+(i\lambda)^2  \sum_{\omega,\omega'} \sum_{i,j} J^{ij}(\omega)[
	S^{i^{\dagger}}_{\omega} \, S^{j}_{\omega'}O_{S}(t) -S^{i^{\dagger}}_{\omega}O_{S}(t) S^{j}_{\omega'}
	]+ h.c.
\end{align}  
where $J^{ij}(\omega)$ is the Fourier transformation of the bath correlation functions and $S^{i}_{\omega}$ is the Fourier decomposition $\tilde{S^i}(t)$ and are given as, 
\begin{align}
	\label{eq100a}
	J^{ij}(\omega)&=\int_{0}^{\infty}d\tau e^{-i \omega \tau} \, {\rm Tr}_{R} [\Tilde{R}^i (0)\Tilde{R}^j(-\tau)\rho_{R}] \nonumber\\
	\Tilde{S^i}(t)&=\sum_{\omega} S^i_{\omega} e^{i \omega t}
\end{align}

One can simplify further the expressions for the  $I$ in the Markovian limit. Let us first analyse Eq.~\eqref{eqI4} in this limit.
As in the Markovian limit the bath correlation functions i.e., ${\rm Tr}_{R}[\Tilde{R}(-\tau_{1})\Tilde{R}(-\tau_{2})\rho_{R}$ is a rapidly decaying function of $\tau_{2}-\tau_{1}$ only, this implies, we can ignore the bath correlation after a characteristic time scale $\tau_{B}$, determined by bath dynamics. Using this fact, it becomes easy to analyse equation \eqref{eqI4}. Let us first  calculate the $\tau_{2}$ integration by assuming $t_{2}>t_{1}$.\footnote{With out loss of generality, we use   $t_{2}>t_{1}$ for rest of the discussion.} Significant contribution to the integral will come with in the range $|\tau_2-\tau_1|\le \tau_B$ which gives $\tau_1-\tau_B\le\tau_2 \le \tau_B+\tau_1.$ Using this fact, we can write equation \eqref{eqI4} as
\begin{widetext}
\begin{align}\label{simI4}
 I_1 &=-\lambda^2\sum_{\omega,\omega'}\sum_{i,j}O_{1  S}(t_{1}) S^i_{\omega} O_{2  S}(t_{2})S^j_{\omega'}\int^{t_{1} }_{ 0}d\tau_{1}e^{-i\omega\tau_{1}} \int_{\tau_{1}-\tau_{B}}^{\tau_{1}+\tau_{B}} d\tau_{2} e^{-i\omega'\tau_{2}}{\rm Tr}_{R}[\Tilde{R^i}(-\tau_{1})\Tilde{R^j}(-\tau_{2})\rho_{R}] \nonumber\\
 &=-\lambda^2\sum_{\omega,\omega'}\sum_{i,j}O_{1  S}(t_{1}) S^i_{\omega} O_{2  S}(t_{2})S^j_{\omega'}\int^{t_{1} }_{ 0}d\tau_{1}e^{-i\omega\tau_{1}} \int_{-\tau_{B}}^{\tau_{B}} d\tau e^{-i\omega'(\tau_{1}-\tau)}{\rm Tr}_{R}[\Tilde{R^i}(0)\Tilde{R^j}(\tau)\rho_{R}]
 \end{align}
 \end{widetext}
 where in the last line we have used the variable $\tau=\tau_{1}-\tau_{2}$ to rewrite the integral.
In the Markovian limit, the correlation function decays very fast beyond $\tau_B,$ this implies that the integration limit in the last line of \eqref{simI4} can be extended to infinity. This gives  \begin{widetext}
\begin{align}\label{eq26'}
 I_1 
 &=-\lambda^2\sum_{\omega,\omega'}\sum_{i,j}O_{1  S}(t_{1}) S^i_{\omega} O_{2  S}(t_{2})S^j_{\omega'}\int_{0}^{t_{1}}d\tau_{1}e^{-i(\omega+\omega')\tau_{1}} \int_{-\infty}^{\infty} d\tau e^{i\omega'\tau} \, {\rm Tr}_{R} [\Tilde{R}^i(0)\Tilde{R}^j(\tau)\rho_{R}] \, \nonumber\\
 &= -\lambda^2\sum_{\omega,\omega'}\sum_{i,j}O_{1  S}(t_{1}) S^i_{\omega} O_{2  S}(t_{2})S^j_{\omega'} \alpha_{1}(\omega,\omega',t_{1}) \beta_{1}^{ij}(\omega')
 \end{align}
 \end{widetext}
where in the last line we have defined $\alpha_1$ and $\beta_1$ for simplicity and their explicit expression is given below \eqref{eq28'}.
By following the identical steps we can find $I_{2},I_{3}$ and $I_{4}$

\bigskip
\begin{align} \label{eq27'}
 I_{2} = &\lambda^2   \sum_{\omega,\omega'} \sum_{i,j}S^{i \dagger}_{  \omega}O_{1  S}(t_{1})
   O_{2  S}(t_{2})
  S^j_{\omega'}\; \alpha_{2}(\omega,\omega',t_{1}) \;\beta_{2}^{ij}(\omega'),\nonumber\\
 I_{3} = &\lambda^2   \sum_{\omega,\omega'} \sum_{i,j} O_{1  S}(t_{1})
  S^i_{\omega} S^{j \dagger}_{\omega'} O_{2  S}(t_{2})
  \; \alpha_{3}(\omega,\omega',t_{1}) \;\beta_{3}^{ij}(\omega'), \nonumber \\
   I_{4} = &-\lambda^2   \sum_{\omega,\omega'} \sum_{i,j} S^{i \dagger}_{  \omega} O_{1  S}(t_{1})
 S^{j \dagger}_{  \omega'}O_{2  S}(t_{2})
  \; \alpha_{4}(\omega,\omega',t_{1}) \;\beta_{4}^{ij}(\omega').
\end{align}
where,
\begin{widetext}
\begin{equation} \label{eq28'}
\begin{split}
    &\alpha_{1}(\omega,\omega',t_{1})=\int_{0}^{t_{1}}d\tau_{1}e^{-i(\omega+\omega')\tau_{1}}\;\;\;\;\;\;\;\;\;\; \beta_{1}^{ij}(\omega')=\int_{-\infty}^{\infty}d\tau e^{i \omega'\tau} \, {\rm Tr}_{R} [\Tilde{R}  ^{i } (0)\Tilde{R}  ^{j }(\tau)\rho_{R}] \\
    &  \alpha_{2}(\omega,\omega',t_{1})=\int_{0}^{t_{1}}d\tau_{1}e^{i(\omega-\omega')\tau_{1}} \;\;\;\;\;\;\;\;\;\; \beta_{3}^{ij}(\omega')=\int_{-\infty}^{\infty}d\tau e^{-i\omega'\tau}  \, {\rm Tr}_{R} [\Tilde{R}  ^{i } (0)\Tilde{R}^j(\tau)\rho_{R}]\\
    &\alpha_{3}(\omega,\omega',t_{1})=\int_{0}^{t_{1}}d\tau_{1}e^{-i(\omega-\omega')\tau_{1}} \;\;\;\;\;\;\;\;\;   \beta_{2}^{ij}(\omega')=\beta_{1}^{ij}(\omega') \\
    & \alpha_{4}(\omega,\omega',t_{1})=\int_{0}^{t_{1}}d\tau_{1}e^{i(\omega+\omega')\tau_{1}}  \;\;\;\;\;\;\;\;\;  \beta_{4}^{ij}(\omega')=\beta_{3}^{ij}(\omega')
\end{split}
\end{equation}
\end{widetext}
Using equation \eqref{eq26'}, \eqref{eq27'} we  get the explicit form of the irreducible part $I$ (\eqref{eq2}) in the Markovian limit.

We can further simplify the expression of $\alpha$'s defined in equation \eqref{eq28'} using the secular approximation. Let us first substitute, $s_{1}=\lambda^2 \tau_{1}$ and $\sigma=\lambda^2 t_{1}$ in the expression of $\alpha_{1}$ define in \eqref{eq28'},
\begin{equation}\label{eqa00}
    \alpha_{1}(\omega,\omega',t_{1})= \lambda^2\int_{0}^{t_{1}}d\tau_{1}e^{i(\omega+\omega')\tau_{1}}=\lim_{\lambda \to 0}\int_{0}^{\sigma}ds_{1}e^{i\frac{(\omega+\omega')}{\lambda^2}s_{1}}
\end{equation}
Now the Riemann-Lebesgue lemma states that,
if $f(t)$ is an integrable function in $[a, b]$ then
\begin{equation}\label{eqrll}
  \lim_{x \to \infty} \int_{a}^{b}dt e^{ixt}f(t)=0.
\end{equation}
By comparing the equation \eqref{eqa00} with the Riemann-Lebesgue lemma \eqref{eqrll}, we can conclude that $ \alpha_{1}$ is non-zero only when $(\omega+\omega')=0$ i.e. $ \alpha_{1}(\omega,\omega',t_{1})=\lambda^2 t_{1}\delta_{\omega,-\omega'}$ . 
Similarly, we can show that, $ \alpha_{4}(\omega,\omega',t_{1})= \alpha_{1}(\omega,\omega',t_{1})=\lambda^2 t_{1}\delta_{\omega,-\omega'}$ and $ \alpha_{2}(\omega,\omega',t_{1})= \alpha_{3}(\omega,\omega',t_{1})=\lambda^2 t_{1}\delta_{\omega,\omega'}$.\\

In the Markovian-secular limit, the master equation for the one-point reduced operator Eq.~\eqref{eq14a} further simplifies to,  
\begin{align}\label{eq14a0}
	&\frac{d}{dt}O_{S}(t)=i H_{S} O_{S}(t)\nonumber\\
	&+(i\lambda)^2  \sum_{\omega} \sum_{i,j} J^{ij}(\omega)[
	S^{i^{\dagger}}_{\omega} \, S^{j}_{\omega}O_{S}(t) -S^{i^{\dagger}}_{\omega}O_{S}(t) S^{j}_{\omega}
	]+ h.c.
\end{align}  
This master equation is used to derive Eq.~(\ref{eql}) for the dissipative spin-boson model in the main text.

\section{Some further details on QRT in the Markovian limit}
In this Appendix, we discuss some further details on QRT in the Markovian limit and in particular some useful properties of the function $I$ as also discussed in the main text. One of the interesting properties of the function $I$ is to note that its dependence on $t_2$ is quite simple. This in turn helps us to establish Regression theorem quite easily. 

\subsection{Irreducible part for two-point function}\label{tptra}
Let us first set, $O_{2}=A_{\mu}$ and
 assume $t_{1}<t_{2}$ in equation \eqref{eq2}. Taking derivative of the equation \eqref{eq2} with respect to $t_{2},$ 1st term of the right hand side of the equation \eqref{eq2} trivially gives the regression type form  (using equation \eqref{eqe} ). Equations \eqref{eqf}, \eqref{eq26'}-\eqref{eq28'} gives us the explicit form of the 2nd term i.e. irreducible part of the equation \eqref{eq2}. Notice that, the $t_{2}$ dependency in the irreducible part  $I[O_{S}(t_{1}),A_{\mu  S}(t_{2})]$ comes from $A_{\mu  S}(t_{2})$ only. So if we take derivative of $I$ with respect to $t_{2}$, it simply gives
\begin{equation}
    \frac{d}{dt_{2}}\label{eqg'} I[O_{S}(t_{1}),A_{\mu  S}(t_{2})] =\sum_{\lambda} M_{\mu \lambda} I[O_{S}(t_{1}),A_{\lambda  S}(t_{2})]
\end{equation}
In addition, in the expression of $I[O_{S}(t_{1}),A_{\mu  S}(t_{2})]$ (see equation \eqref{eq26'}-\eqref{eq28'}  ), it is easy to notice that if we swap the position of $O_{S}(t_{1})$ and $A_{\mu  S}(t_{2})$, the equation \eqref{eqg'} still holds i.e.
\begin{equation}
    \frac{d}{dt_{2}} I[A_{\mu  S}(t_{2}),O_{S}(t_{1})] =\sum_{\lambda} M_{\mu \lambda} I[A_{\lambda  S}(t_{2}),O_{S}(t_{1})]
\end{equation}

Interestingly, If we consider $t_{2}<t_{1}$ and take derivative of the equation \eqref{eq2} with respect to $t_{2}$, then we will not get the regression type form . The simple reason behind this in Heisenberg picture is that in the expression of $I$, $t_{2}$ dependency comes from both $A_{\mu  S}$ and $\alpha$'s defined in \eqref{eq28'}. Now taking derivative with respect to $t_{2}$ will give rise to complicated terms\footnote{We shall see that this observation is true for general multi time correlation function.  In Schrodinger picture as well, we have seen that regression theorem holds only for derivatives with respect to highest time. } i.e.
\begin{equation}\label{eq102a}
    \frac{d}{dt_{2}} \langle O(t_{1}) A_{\mu}(t_{2})\rangle \neq \sum_{\lambda} M_{\mu \lambda} \langle O(t_{1}) A_{\lambda}(t_{2})\rangle
\end{equation}

\subsection{Irreducible part for three-point function}\label{Tptra}

\label{threereg}
Now let us first set, $O_{3}=A_{\mu}$ and
 assume $t_{i}<t_{3}$ with $i=1,2$ in equation \eqref{eq22}. Taking  derivative of the equation \eqref{eq22} with respect to $t_{3},$ 1st term of the right hand side gives (using equation \eqref{eqe} ),
\begin{align}\label{eqMa}
   & \frac{d}{dt_{3}} (O_{1  S}(t_{1})O_{2  S}(t_{2})A_{\mu  S}(t_{3}))\nonumber\\
   &=\sum_{\lambda} M_{\mu \lambda} (O_{1  S}(t_{1})O_{2  S}(t_{2})A_{\lambda  S}(t_{3}))
\end{align}
Now the 3rd term of the right hand side of the equation \eqref{eq22} is (using equation \eqref{eqf},
\begin{align} \label{eqNa}
    & W_{1,2,3}\Big\{O_{1  S}(t_{1})I[O_{2  S}(t_{2}),A_{\mu  S}(t_{3})]\Big\} \nonumber\\
    &=W_{1,2,3}\Big\{O_{1  S}(t_{1})( I_{1}+I_{2}+I_{3}+I_{4})\Big\} \end{align}
The first term of the right hand side of the equation \eqref{eqNa} in the markovian limit is given by (using equation \eqref{eq27'}),
\begin{widetext}
\begin{align} \label{eq277'}
  W_{1,2,3}\Big\{O_{1  S}(t_{1})I_{1}[O_{2  S}(t_{2}),A_{\mu  S}(t_{3})]\Big\}=
   -\lambda^2   \sum_{\omega,\omega'} \sum_{i,j} O_{1 S}(t_{1}) O_{2  S}(t_{2})S^{i }_{  \omega}
 A_{\mu S}(t_{3})S^{j }_{  \omega'}
  \; \alpha_{1}(\omega,\omega',t_{2}) \;\beta_{1}^{ij}(\omega')
\end{align}
\end{widetext}
If we differentiate the equation \eqref{eq277'} with respect to $t_{3}$, we get (using equation \eqref{eqe} ) 
\begin{align} \label{eq288'}
    &\frac{d}{dt_{3}}  W_{1,2,3}\Big\{O_{1  S}(t_{1})I_{1}[O_{2  S}(t_{2}),A_{\mu  S}(t_{3})]\Big\}\nonumber\\
    &=\sum_{\lambda} M_{\mu \lambda}    W_{1,2,3}\Big\{O_{1  S}(t_{1})I_{1}[O_{2  S}(t_{2}),A_{\lambda  S}(t_{3})]\Big\}
\end{align}
since in the expression of $W_{1,2,3}\Big\{O_{1  S}(t_{1})I_{1}[O_{2  S}(t_{2}),A_{\mu  S}(t_{3})]\Big\}$, $t_{3}$ dependency comes from $A_{\mu  S}(t_{3})$ only. Similarly, we can show that all the other terms of the equation \eqref{eqNa} follows the identical equation to \eqref{eq288'}. This finally gives,
\begin{align} \label{eqN'}
    &\frac{d}{dt_{3}}  W_{1,2,3}\Big\{O_{1  S}(t_{1})I[O_{2  S}(t_{2}),A_{\mu  S}(t_{3})]\Big\} \nonumber\\
    &=\sum_{\lambda} M_{\mu \lambda}  W_{1,2,3} \Big\{O_{1  S}(t_{1})I[O_{2  S}(t_{2}),A_{\lambda S}(t_{3})]\Big\}
\end{align}

\subsection{Irreducible part for four-point function}\label{fptra}

Set, $   O_{4}=A_{\mu}$, $t_{i}<t_{4}$ with $i=1,2,3$ in the equation \eqref{eq88} and if we take derivative of the equation \eqref{eq88} with respect to $t_{4}$ then the 1st term of the right hand side will simply give (using equation \eqref{eqe} ), 
    \begin{align}
   & \frac{d}{dt_{4}} (O_{1  S}(t_{1})O_{2  S}(t_{2})O_{3  S}(t_{3})A_{\mu  S}(t_{4})) \nonumber\\
    &=\sum_{\lambda} M_{\mu \lambda} (O_{1  S}(t_{1})O_{2  S}(t_{2})O_{3  S}(t_{3})A_{\lambda  S}(t_{4}))
\end{align}
  We can very straight forwardly conclude that the 2nd term will obey the following equation (using equation \eqref{eqe}),
  \begin{align}\label{eq44}
    &\frac{d}{dt_{4}} W_{1,2,3,4}\Big\{I[O_{1  S}(t_{1}),O_{2  S}(t_{2})]O_{3  S}(t_{3})A_{\mu  S}(t_{4})\Big\} \nonumber\\
    &=\sum_{\lambda} M_{\mu \lambda} W_{1,2,3,4}\Big\{I[O_{1  S}(t_{1}),O_{2  S}(t_{2})]O_{3  S}(t_{3})A_{\lambda  S}(t_{4})\Big\}
\end{align}
   By giving the exactly similar argument we can show that the 3rd and 4th term of the right hand side of the equation \eqref{eq88} will follow the identical to the above equation. Now using equation \eqref{eqg'} we can show that the 5th term of the right hand side of the same equation will give,
    \begin{align}
   & \frac{d}{dt_{4}} W_{1,2,3,4}\Big\{I[A_{\mu  S}(t_{4}),O_{3  S}(t_{3})]O_{2  S}(t_{2})O_{1  S}(t_{1})\Big\}\nonumber\\
   &=\sum_{\lambda} M_{\mu \lambda} W_{1,2,3,4}\Big\{I[A_{\lambda  S}(t_{4}),O_{3  S}(t_{3})]O_{2  S}(t_{2})O_{1  S}(t_{1})\Big\}
\end{align}
  Since, in the expression of $I[A_{\mu  S}(t_{4}),O_{3  S}(t_{3})]$,  the $t_{4}$ dependency comes from $A_{\mu  S}(t_{2})$ only and by giving the similar argument we can show that the last two term of the equation  \eqref{eq88} will also follow the identical equation. Then finally we will arrive at the following equation (using equation \eqref{eq88} ),
  \begin{align}\label{eq46a}
   & \frac{d}{dt_{4}} (O_{1}(t_{1})O_{2}(t_{2})O_{3}(t_{3})A_{\mu}(t_{4}))_{S} \nonumber\\
   &=\sum_{\lambda} M_{\mu \lambda} (O_{1}(t_{1})O_{2}(t_{2})O_{3}(t_{3})A_{\lambda}(t_{4}))_{S}
\end{align}


\section{ Out-of-time-ordered correlators (OTOCs)}
\label{otoc}
In this section we calculate out-of-time-order correlator (OTOC). More specifically we want to derive regression type theorem for OTOC \cite{otoc1}. Now set, $t_{1}=t_{3},t_{2}=t_{4},t_{2}>t_{1}$ and $O_{2}(t_{2})=A_{\mu}(t_{2})$ , $O_{4}(t_{2})=A_{\nu}(t_{2})$ in the equation \eqref{eq88} then we get,
\begin{widetext}
   \begin{equation} \label{eq 3}
\begin{split}
(O_{1}(t_{1})A_{\mu}(t_{2})O_{3}(t_{1})A_{\nu}(t_{2}))_{S}&=O_{1  S}(t_{1})A_{\mu  S}(t_{2})O_{3  S}(t_{1})A_{\nu  S}(t_{2}) +W_{1,2,3,4}\Big \{I[O_{1  S}(t_{1}),A_{\mu  S}(t_{2})] O_{3  S}(t_{1})A_{\nu  S}(t_{2})\Big\} \\
     & +W_{1,2,3,4} \Big\{O_{1  S}(t_{1}) I[A_{\mu  S}(t_{2}),O_{3  S}(t_{1})] A_{\nu  S}(t_{2})\Big\} +W_{1,2,3,4} \Big\{I[O_{1  S}(t_{1}), O_{3  S}(t_{1})]A_{\mu  S}(t_{2})A_{\nu  S}(t_{2})\Big\}\\
     &+W_{1,2,3,4} \Big\{I[A_{\nu  S}(t_{2}), O_{3  S}(t_{1})]A_{\mu  S}(t_{2})O_{1  S}(t_{1})\Big\} 
     W_{1,2,3,4} \Big\{I[O_{1  S}(t_{1}),A_{\nu  S}(t_{2})] O_{3  S}(t_{1})A_{\mu  S}(t_{2})\Big\} \\
     & +W_{1,2,3,4} \Big\{O_{1  S}(t_{1}) I[A_{\mu  S}(t_{2}),A_{\nu  S}(t_{2})] O_{3  S}(t_{1})\Big\} 
\end{split}
\end{equation}
\end{widetext}
Differentiate the above equation by $t_{2}$,
  then the first term of the right hand side gives
  \begin{equation} \label{eq139}
\begin{split}
& \frac{d}{dt_{2}} (O_{1  S}(t_{1})A_{\mu  S}(t_{2})O_{3  S}(t_{1})A_{\nu  S}(t_{2}))\nonumber\\
&=\sum_{\lambda} M_{\mu \lambda}(O_{1  S}(t_{1})A_{\lambda  S}(t_{2})O_{3  S}(t_{1})A_{\nu  S}(t_{2})) \\
     & +\sum_{\lambda'} M_{\nu \lambda'}(O_{1  S}(t_{1})A_{\mu  S}(t_{2})O_{3  S}(t_{1})A_{\lambda'  S}(t_{2}))
\end{split}
\end{equation}
Similarly, the 2nd term of the right hand side gives,
   \begin{equation} \label{eq140}
\begin{split}
 & \frac{d}{dt_{2}} W_{1,2,3,4} \Big\{I[O_{1  S}(t_{1}),A_{\mu  S}(t_{2})] O_{3  S}(t_{1})A_{\nu  S}(t_{2})\Big\}\\
 &=\sum_{\lambda} M_{\mu \lambda}W_{1,2,3,4} \Big\{I[O_{1  S}(t_{1}),A_{\lambda  S}(t_{2})] O_{3  S}(t_{1})A_{\nu  S}(t_{2})\Big\} \\
     & +\sum_{\lambda'} M_{\nu \lambda'}W_{1,2,3,4} \Big\{I[O_{1  S}(t_{1}),A_{\mu  S}(t_{2})] O_{3  S}(t_{1})A_{\lambda'  S}(t_{2})\Big\}
\end{split}
\end{equation}
 All the other terms in \eqref{eq 3} also follow the identical equation except the last term. The last term gives,
 \begin{equation} \label{eqp}
\begin{split}
 & \frac{d}{dt_{2}} W_{1,2,3,4} \Big\{O_{1  S}(t_{1}) I[A_{\mu  S}(t_{2}),A_{\nu  S}(t_{2})] O_{3  S}(t_{1})\Big\}\\
 &=W_{1,2,3,4} \Big\{O_{1  S}(t_{1})  \frac{d}{dt_{2}}(I[A_{\mu  S}(t_{2}),A_{\nu  S}(t_{2})]) O_{3  S}(t_{1})\Big\} 
\end{split}
\end{equation}
 Using expresssion of $I$
   \begin{equation} \label{eq142}
\begin{split}
 \frac{d}{dt_{2}}I[A_{\mu  S}(t_{2}),A_{\nu  S}(t_{2})]& =\sum_{\lambda} M_{\mu \lambda}I[A_{\lambda  S}(t_{2}),A_{\nu  S}(t_{2})] \\
 &+\sum_{\lambda'} M_{\nu \lambda'}I[A_{\mu  S}(t_{2}),A_{\lambda'  S}(t_{2})])\\
 &+F[A_{\mu  S}(t_{2}),A_{\nu  S}(t_{2})]
\end{split}
\end{equation}
  where
    \begin{equation} \label{eq143}
\begin{split}
&F[A_{\mu  S}(t_{2}),A_{\nu  S}(t_{2})] \\
&=-\lambda^2   \sum_{\omega,\omega'} \sum_{i,j} S^{i \dagger}_{  \omega} A_{\mu  S}(t_{2})
 S^{j \dagger}_{  \omega'}A_{\nu  S}(t_{2})
  \; e^{i(\omega+\omega')t_{2}} \;\beta_{4}^{ij}(\omega')\\
  &+\lambda^2   \sum_{\omega,\omega'} \sum_{i,j} S  ^{i \dagger}_{\omega}A_{\mu  S}(t_{2})
   A_{\nu  S}(t_{2})
  S^j_{\omega'}\; e^{i(\omega-\omega')t_{2}} \;\beta_{2}^{ij}(\omega')\\
  &+\lambda^2   \sum_{\omega,\omega'} \sum_{i,j} A_{\mu  S}(t_{2})
  S^i_{\omega} S^{j \dagger}_{\omega'} A_{\nu  S}(t_{2})
  \; e^{-i(\omega-\omega')t_{2}} \;\beta_{3}^{ij}(\omega')\\
  &-\lambda^2   \sum_{\omega,\omega'} \sum_{i,j} A_{\mu  S}(t_{2})
  S^i_{\omega} A_{\nu  S}(t_{2})
  S^j_{\omega'}\; e^{-i(\omega+\omega')t_{2}} \;\beta_{1}^{ij}(\omega')
\end{split}
\end{equation}
 \eqref{eqp} becomes
 \begin{equation} \label{eq144}
\begin{split}
  &\frac{d}{dt_{2}} W_{1,2,3,4} \Big\{O_{1  S}(t_{1}) I[A_{\mu  S}(t_{2}),A_{\nu  S}(t_{2})] O_{3  S}(t_{1})\Big\} \\
  &=\sum_{\lambda} M_{\mu \lambda}W_{1,2,3,4} \Big\{O_{1  S}(t_{1})  I[A_{\lambda  S}(t_{2}),A_{\nu  S}(t_{2})] O_{3  S}(t_{1})\Big\} \\
  &+\sum_{\lambda'} M_{\nu \lambda'}W_{1,2,3,4} \Big\{O_{1  S}(t_{1})  I[A_{\mu  S}(t_{2}),A_{\lambda'  S}(t_{2})] O_{3  S}(t_{1})\Big\} \\
  &+W_{1,2,3,4}\Big \{O_{1  S}(t_{1})   F[A_{\mu  S}(t_{2}),A_{\nu  S}(t_{2})]O_{3  S}(t_{1})\Big\} 
\end{split}
\end{equation}
 Finally adding all the different contributions we receive,
\begin{equation} \label{eq 2'}
\begin{split}
 & \frac{d}{dt_{2}}(O_{1}(t_{1})A_{\mu}(t_{2})O_{3}(t_{1})A_{\nu}(t_{2}))_{S}\\
 &=\sum_{\lambda} M_{\mu \lambda}(O_{1}(t_{1})A_{\lambda}(t_{2})O_{3}(t_{1})A_{\nu}(t_{2}))_{S}\\
  &+\sum_{\lambda'} M_{\nu \lambda'}(O_{1}(t_{1})A_{\mu}(t_{2})O_{3}(t_{1})A_{\lambda'}(t_{2}))_{S} \\
  &+W_{1,2,3,4} \Big\{O_{1  S}(t_{1})   F[A_{\mu  S}(t_{2}),A_{\nu  S}(t_{2})]O_{3  S}(t_{1})\Big\} 
\end{split}
\end{equation}\\
We observe that regression theorem for OTOC takes a different form than discussed in previous section.\\
\\
\\

\section{Example: Dissipative spin half system }
\label{tpds0}
In this appendix we provide the details of the calculation for the dissipative spin 1/2 system as discussed in the main text. 
\subsection{Correlation function in Heisenberg picture} 
 We give out expressions for two, three and four-point reduced operators using the Heisenberg picture.

\subsubsection{{Two point function}}\label{tpds}
Here we are going to calculate the irreducible part of the equation \eqref{eq2} to get the following two-point functions $     \langle\sigma_{x}(t)\sigma_{x}(t+\tau)\rangle$ and $\langle\sigma_{x}(t)\sigma_{y}(t+\tau)\rangle$. Using equation
\eqref{eqf}, \eqref{eq26'}-\eqref{eq28'} and by imposing secular approximation we can calculate the irreducible part $I$ and it is given by,
\textcolor{black}
{\begin{equation} \label{eqIex02}
\begin{split}
  I[ \sigma_{x  S}(t_{1}), \sigma_{x  S}(t_{2})] = \gamma t_{1} e^{i\omega'_{0}(t_{2}-t_{1})}\begin{pmatrix}  1 & 0  \\ 0 & 1 \end{pmatrix}
\end{split}
\end{equation}\\
\begin{equation} \label{eqIex20}
\begin{split}
   \sigma_{x  S}(t_{1}) \sigma_{x  S}(t_{2}) =\left(1- \frac{\gamma}{2} (t_{1}+t_{2})\right)\begin{pmatrix}  e^{-i\omega'_{0}(t_{2}-t_{1})} & 0  \\ 0 & e^{i\omega'_{0}(t_{2}-t_{1})} \end{pmatrix}
\end{split}
\end{equation}
\begin{equation} \label{eqIex2}
\begin{split}
  I[ \sigma_{x  S}(t_{1}), \sigma_{y  S}(t_{2})] = -i\gamma t_{1} e^{i\omega'_{0}(t_{2}-t_{1})}\begin{pmatrix}  1 & 0  \\ 0 & 1 \end{pmatrix}
\end{split}
\end{equation}
\begin{equation} \label{eqIex20a}
\begin{split}
   \sigma_{x  S}(t_{1}) \sigma_{y  S}(t_{2}) =\left(1- \frac{\gamma}{2} (t_{1}+t_{2})\right)\begin{pmatrix}  ie^{-i\omega'_{0}(t_{2}-t_{1})} & 0  \\ 0 & -ie^{i\omega'_{0}(t_{2}-t_{1})} \end{pmatrix}
\end{split}
\end{equation}
}
\subsubsection{Three and four-point functions}
To get three and four point reduced operators we have to calculate all the $W$ terms of the equation \eqref{eq22} and equation \eqref{eq88}. Here we have written down the expression of all the $W$ terms defined in equation \eqref{eq22},
\\
\textcolor{black}
{\begin{widetext}
\begin{equation} \label{eqwex02}
\begin{split}
  W_{1,2,3} \{I[ \sigma_{x  S}(t_{1}), \sigma_{x S}(t_{2})] \sigma_{x  S}(t_{3})\} =  \gamma t_{1} \begin{pmatrix}  0 & e^{i\omega'_{0}(-t_{1}+t_{2}+t_{3})}  \\  e^{i\omega'_{0}(-t_{1}+t_{2}-t_{3})} & 0\end{pmatrix}
\end{split}
\end{equation}
\begin{equation} \label{eqwex2}
\begin{split}
 W_{1,2,3} \{I[ \sigma_{x  S}(t_{1}), \sigma_{x  S}(t_{3})] \sigma_{x  S}(t_{2})\}=  -\gamma t_{1} \begin{pmatrix}  0 & e^{i\omega'_{0}(-t_{1}+t_{2}+t_{3})}  \\  e^{i\omega'_{0}(-t_{1}-t_{2}+t_{3})} & 0\end{pmatrix}
\end{split}
\end{equation}
\begin{equation} \label{eqwex20}
\begin{split}
  W_{1,2,3} \{\sigma_{x S}(t_{1})I[ \sigma_{x S}(t_{2}), \sigma_{x  S}(t_{3})] \} =  \gamma t_{2} \begin{pmatrix}  0 & e^{i\omega'_{0}(t_{1}-t_{2}+t_{3})}  \\  e^{i\omega'_{0}(-t_{1}-t_{2}+t_{3})} & 0\end{pmatrix}
\end{split}
\end{equation}
and
\begin{equation} \label{eqIex200}
\begin{split}
   \sigma_{x  S}(t_{1}) \sigma_{x S}(t_{2})\sigma_{x S}(t_{3})  =\left(1- \frac{\gamma}{2} (t_{1}+t_{2}+t_{3})\right)\begin{pmatrix}  0&e^{i\omega'_{0}(t_{1}-t_{2}+t_{3})}  \\  e^{i\omega'_{0}(-t_{1}+t_{2}-t_{3})}&0 \end{pmatrix}
\end{split}
\end{equation}
Here we have shown only one $W$ term which is nedded to calculate four-point reduced operator and as defined in equation \eqref{eq88}
\begin{equation} \label{eqwex4}
\begin{split}
  W_{1,2,3,4} \{I[ \sigma_{x S}(t_{1}), \sigma_{x  S}(t_{2})] \sigma_{x  S}(t_{3}) \sigma_{x S}(t_{4})\} = \gamma t_{1} \begin{pmatrix}   e^{i\omega'_{0}(-t_{1}+t_{2}+t_{3}-t_{4})} &0 \\  0&e^{i\omega'_{0}(-t_{1}+t_{2}-t_{3}+t_{4})}
  \end{pmatrix}
\end{split}
\end{equation}
\end{widetext}}

\section{Extension of regression theorem for two-point reduced operator}\label{nmvqrt}
 The two-point reduced operator can be written as,
  \begin{equation}
    ( O_{1}(t_{1})O_{2}(t_{2}))_{S}=O_{1  S}(t_{1})O_{2  S}(t_{2})+
 I[O_{1  S}(t_{1}),O_{2  S}(t_{2})] 
  \end{equation}
Where, $  I=I_{1}+I_{2}+I_{3}+I_{4}$ and we can show that (here we assumed $t_{1}>t_{2}$), (correct upto order $\lambda^2$)
\newpage
\begin{align} \label{eq t}
&I_{1} \nonumber\\& = - \lambda^2\int_{0}^{t_{1}} \int_{0}^{t_{2}}d\tau_{1}d\tau_{2}\; \alpha(\tau_{2}-\tau_{1})O_{1  S}(t_{1})
 \Tilde{ L}^\dagger(-\tau_{1}) O_{2  S}(t_{2})
   \Tilde{L}(-\tau_{2}),\nonumber\\
&I_{2}\nonumber\\& =  \lambda^2\int_{0}^{t_{1}} \int_{0}^{t_{2}}d\tau_{1}d\tau_{2}\; \alpha(\tau_{2}-\tau_{1}) \Tilde{ L}^\dagger(-\tau_{1})O_{1  S}(t_{1})
 O_{2  S}(t_{2})
   \Tilde{L}(-\tau_{2}),\nonumber\\
  & I_{3} \nonumber\\& =  \lambda^2 \int_{0}^{t_{1}} \int_{0}^{t_{2}}d\tau_{1}d\tau_{2}\; \alpha(\tau_{2}-\tau_{1})O_{1  S}(t_{1})
 \Tilde{ L}^\dagger(-\tau_{1})\Tilde{L}(-\tau_{2}) O_{2  S}(t_{2}),
   \nonumber\\
&  I_{4}\nonumber\\ &= - \lambda^2 \int_{0}^{t_{1}} \int_{0}^{t_{2}}d\tau_{1}d\tau_{2}\; \alpha(\tau_{2}-\tau_{1})\Tilde{ L}^\dagger(-\tau_{1})O_{1  S}(t_{1})
  \Tilde{L}(-\tau_{2}) O_{2  S}(t_{2}).
\end{align}
Now by differentiating the equation \eqref{eq t} with respect to $t_{1}$ we receive,
\begin{widetext}
\begin{align} \label{eq u}
 & \frac{d}{dt_{1}} I_{1} =\nonumber\\  &- \lambda^2 \int_{0}^{t_{1}} \int_{0}^{t_{2}}d\tau_{1}d\tau_{2}\; \alpha(\tau_{2}-\tau_{1})(i [H_{S}, O_{1  S}(t_{1})]
 \Tilde{ L}^\dagger(-\tau_{1}) O_{2  S}(t_{2})
   \Tilde{L}(-\tau_{2}))-    \lambda^2\int_{0}^{t_{2}}d\tau_{2}\; \alpha(\tau_{2}-t_{1})O_{1  S}(t_{1})
 \Tilde{ L}^\dagger(-t_{1}) O_{2  S}(t_{2})
   \Tilde{L}(-\tau_{2})\nonumber \\ 
 &\frac{d}{dt_{1}} I_{2} =\nonumber\\  & \lambda^2\int_{0}^{t_{1}} \int_{0}^{t_{2}}d\tau_{1}d\tau_{2}\; \alpha(\tau_{2}-\tau_{1})( \Tilde{ L}^\dagger(-\tau_{1}) [iH_{S}, O_{1  S}(t_{1})]
 O_{2  S}(t_{2})
   \Tilde{L}(-\tau_{2}))+ \lambda^2\int_{0}^{t_{2}}d\tau_{2}\; \alpha(\tau_{2}-t_{1}) \Tilde{ L}^\dagger(-t_{1})O_{1  S}(t_{1})
 O_{2  S}(t_{2})
   \Tilde{L}(-\tau_{2})\nonumber \\
& \frac{d}{dt_{1}} I_{3}\nonumber=\\   &  \lambda^2\int_{0}^{t_{1}} \int_{0}^{t_{2}}d\tau_{1}d\tau_{2}\; \alpha(\tau_{2}-\tau_{1})(i [H_{S}, O_{1  S}(t_{1})]
 \Tilde{ L}^\dagger(-\tau_{1}) \Tilde{L}(-\tau_{2}) O_{2  S}(t_{2})
  ) + \lambda^2\int_{0}^{t_{2}}d\tau_{2}\; \alpha(\tau_{2}-t_{1})O_{1  S}(t_{1})
 \Tilde{ L}^\dagger(-t_{1}) \Tilde{L}(-\tau_{2}) O_{2  S}(t_{2})
  \nonumber\\
 & \frac{d}{dt_{1}} I_{4}=\nonumber\\
 &  - \lambda^2\int_{0}^{t_{1}} \int_{0}^{t_{2}}d\tau_{1}d\tau_{2}\; \alpha(\tau_{2}-\tau_{1})(\Tilde{ L}^\dagger(-\tau_{1})[i H_{S}, O_{1  S}(t_{1})] \Tilde{L}(-\tau_{2})
 O_{2  S}(t_{2})
  )  -  \lambda^2 \int_{0}^{t_{2}}d\tau_{2}\; \alpha(\tau_{2}-t_{1}) \Tilde{ L}^\dagger(-t_{1})O_{1  S}(t_{1})
  \Tilde{L}(-\tau_{2})O_{2  S}(t_{2})
\end{align}
\end{widetext}
and,
 \begin{equation}
    \frac{d}{dt_{1}} \Big (O_{1  S}(t_{1})O_{2  S}(t_{2})\Big)=\dot O_{1  S}(t_{1})O_{2  S}(t_{2})
  \end{equation}
  In the Markovian limit, we can show that the 2nd term of the equations \eqref{eq u} vanishes.\\
Now assume there exists a set of system operators such that,
 \begin{equation}\label{eq v}
     \frac{d}{dt} A_{\mu  S}(t)=\sum_{\lambda} M_{\mu  \lambda}(t)A_{\lambda  S}(t)
  \end{equation}
Now set, $  O_{2}=O$ and $ O_{1}=A_{\mu}$ then by using equation \eqref{eq u} and \eqref{eq v} we get,
\begin{equation} \label{eq171}
\begin{split}
 &\frac{d}{dt_{1}} I_{1}[A_{\mu  S}(t_{1}),O_{S}(t_{2})] = \sum_{\lambda} M_{\mu \lambda}(t_{1}) I_{4}[A_{\lambda  S}(t_{1}),O_{S}(t_{2})]\\
   &-  \lambda^2\int_{0}^{t_{2}}d\tau_{2}\; \alpha(\tau_{2}-t_{1})[A_{\mu  S}(t_{1})
 \Tilde{ L}^\dagger(-t_{1}) O_{2  S}(t_{2})
   \Tilde{L}(-\tau_{2})]
\end{split}
\end{equation}
We also get the identical equation for $I_{2},I_{3}$ and $I_{4}$ and then adding all of them we will finally get the equation \eqref{eq1nva}.

\end{document}